\documentclass{article}

\usepackage[T1]{fontenc}
\usepackage[utf8]{inputenc}

\usepackage{geometry}
\usepackage{setspace}

\usepackage[journal=nalefd,manuscript=article]{achemso}
\setkeys{acs}{articletitle=true}

\usepackage{graphicx}
\usepackage{float}
\newfloat{scheme}{htbp}{los}
\floatname{scheme}{Scheme}
\floatname{chart}{Chart}
\newfloat{graph}{htbp}{loh}

\usepackage{chemformula} 
\usepackage[version = 4]{mhchem} 

\usepackage{authblk}
\author[1]{Khalid N. Anindya*}
\author[1]{Hong Guo}
\affil[1]{Department of Physics, McGill University, Montréal, QC H3A 2T8, Canada}

\title{Tunable Topological Phases in an Organic One-Dimensional Mott Chain: Odd-Haldane (S = 1/2) and Haldane (S = 1)}
\date{*Email: khalid.anindya@mcgill.ca}

\usepackage{multicol}

\begin{document}

\maketitle

\begin{abstract}
Establishing symmetry-protected topological (SPT) phases with interactions in chemically realistic systems remains an open challenge. We show that a single, synthetically plausible organic one-dimensional chain, tunable via chemical modification of its radical sites, hosts two such phases: an odd-Haldane phase of a dimerized $S=\tfrac{1}{2}$ Heisenberg chain and a Haldane phase of an $S=1$ chain realized when Hund coupling locks two $S=\tfrac{1}{2}$ spins per monomer into $S=1$. Density-functional theory places the active manifold deep in the Mott regime ($U/t\!\approx\!126$), justifying a spin-only Heisenberg description; a compact $(t,U)\!\to\!J$ mapping then fixes exchange couplings. Exact diagonalization and DMRG reveal a consistent SPT fingerprint across both phases, including a quantized many-body Zak phase, even-degenerate entanglement spectrum, protected edge spins, and characteristic triplon/Haldane features in $S^{+-}(q,\omega)$. Our results identify a chemically programmable molecular platform for interacting SPT physics in one dimension and suggest concrete spectroscopic routes to organic Haldane spin chains for nanoscale quantum devices.
\end{abstract}

\section*{Keywords}

Symmetry-protected topological phases; Haldane phase; Dimerized spin chain; Magnetic Nanographene; Many-body Zak phase


\begin{multicols}{2}

Carbonyl–triphenyl (CTP) derivatives offer a rigid, planar $\pi$-framework
with orthogonal chemical handles (carbonyls, N-heteroatoms) that enable stable
$\pi$-radicals and on-surface covalent coupling. High-spin aza-triangulene
molecules derived from CTP-like precursors have been synthesized and
characterized at the single-molecule level,\cite{WangJACS2022,CalupitanNanoLett2023}
and closely related radical motifs have been stabilized in extended
two-dimensional $\pi$-networks such as Kagome graphene,\cite{PawlakACSNano2025}
establishing the feasibility of embedding robust spins in covalently bonded
lattices. More broadly, on-surface synthesis of open-shell nanographenes and
diradical frameworks now enables exchange pathways, lattice geometry, and
symmetry to be engineered at the atomic scale, realizing designer spin models
in organic $\pi$-systems.\cite{MishraNature2021,ZhaoNatNano2024,HenriquesDesigner2024,HenriquesBeyond2024,Lawrence_5AGNR2020,Friedrich_MetallicGNR2022}
Together with nanographene spin chains that already control chain length,
parity, termination, and access spin excitations
spectroscopically,\cite{MishraNature2021,ZhaoNatNano2024}
these advances motivate a 1D CTP-based polymer whose topological character
and edge multiplets can be tuned \emph{chemically}, providing a realistic route
toward edge-spin qubits and topological spin-transport building
blocks.\cite{ElsePRL2012,BartlettPRL2010,MeiQST2018}

From a band-structure perspective, the two spin-chain realizations we target
are natural interacting analogues of well-known electronic models. The
dimerized spin-$\tfrac12$ ``odd-Haldane'' chain is the spin counterpart of the
topological Su--Schrieffer--Heeger (SSH) polyacetylene chain: strong and weak
bonds alternate, and the topological sector is selected by which bond is cut
at the ends.\cite{Su1979,HeegerRMP1988}
In the SSH model, dimerization of hopping amplitudes produces a quantized Zak
phase and midgap boundary states for one choice of termination; in the
odd-Haldane chain, dimerization of antiferromagnetic exchanges produces a
quantized many-body Zak phase and fractional $S=\tfrac12$ edge spins for the
bond-centred (``odd'') dimerization pattern.\cite{PollmannSPT2012,ZhaoNatNano2024,MishraNature2021}
By contrast, a spin-1 Haldane chain is uniform in real space but topological
in its \emph{entanglement}: it supports non-local string order and
$S=\tfrac12$ edge modes even without explicit dimerization, as captured in the
AKLT valence-bond picture.\cite{AKLT1987,AKLT1988,HaldanePRL1983,PollmannSPT2012}
Here we show that a single, synthetically plausible CTP-based chain realizes both situations within a single chemical
backbone: an SSH-like, bond-centred odd-Haldane phase for a
spin-$\tfrac12$ radical chain, and a site-centred Haldane phase for a
Hund-coupled spin-1 superatom chain.

Interacting symmetry-protected topological (SPT) phases in one dimension are
distinguished by non-local order and boundary excitations that are robust
against local perturbations preserving the protecting
symmetries.\cite{HaldanePRL1983,HaldanePLA1983,AKLT1987,AKLT1988,PollmannSPT2012,denNijs1989,LiHaldane2008}
In the present CTP-based chain, two archetypes emerge from distinct
radicalization patterns on the same molecular scaffold: a single
$\pi$-radical per monomer yields an $S{=}\tfrac12$ lattice with
$J_1\!\ll\!J_2$ (odd-Haldane), whereas two radicals on the same monomer
undergo strong ferromagnetic locking by Hund-coupling ($J_1{<}0$, $|J_1|\!\gg\!J_2$),
forming $S{=}1$ superatoms (Haldane) coupled antiferromagnetically by $J_2$.
Both chains are centrosymmetric, but the inversion center differs, \emph{site-centred} for $S{=}1$ and \emph{bond-centred} for $S{=}\tfrac12$,
which fixes the Zak-phase origin and governs termination-dependent edge
multiplets.\cite{PollmannSPT2012}

Because band-structure proxies do not apply in interacting 1D systems and
finite samples have edges and terminations, we establish SPT order using
experiment-facing many-body diagnostics: a quantized many-body Zak/polarization
phase, an even-degenerate entanglement spectrum at symmetry-respecting cuts,
termination-dependent ground-state multiplets, and real-space edge
magnetization with an exponential envelope.\cite{Zak1989,Resta1998,LiHaldane2008,PollmannES2010,PollmannSPT2012,denNijs1989}
A key advantage of our platform is comparability: both phases arise from the
\emph{same} molecular backbone and protecting symmetries, differing only in
intra-unit exchange, which allows an apples-to-apples assessment within one
chemistry. We focus on two complementary low-energy descriptions. First, the
alternating-exchange Heisenberg (AEH) chain for a radical $S=\tfrac12$
lattice,
\begin{equation}
\label{eq:AEH}
H_{\mathrm{AEH}}
=\sum_{j=1}^{L/2}\Big[J_1\,\mathbf{S}_{2j-1}\!\cdot\!\mathbf{S}_{2j}
\;+\;J_2\,\mathbf{S}_{2j}\!\cdot\!\mathbf{S}_{2j+1}\Big],
\end{equation}
with $J_2>J_1\ge 0$ realizing the odd-Haldane SPT in the strongly dimerized
limit. For open chains, which bond is cut selects the SPT sector:\cite{ZhaoNatNano2024,MishraNature2021}
cutting a weak bond ($J_1$) yields an even-Haldane termination with two
$S=\tfrac12$ edge modes and a near-fourfold ground-state manifold in long
chains; cutting a strong bond ($J_2$) removes free edges (trivial ground
state); and mixed terminations give a twofold manifold with a single edge
mode.\cite{ZhaoNatNano2024} In the bulk, $J_2$ pins singlets and the
elementary excitation is a dispersive one-triplon branch with a cosine-like
shape \cite{Barnes1999,ZhaoNatNano2024}, whose open-boundary folding and intensity ridge have been identified in
atomically precise organic chains.

Second, in the Hund-coupled case two $S=\tfrac12$ on each monomer bind
ferromagnetically ($J_1<0$, $|J_1|\!\gg\!J_2$) into a rigid $S{=}1$ superatom.
Projecting the inter-monomer coupling onto the triplet subspace gives an
\emph{effective} antiferromagnetic $S{=}1$ chain with exchange set by the
inter-monomer bridges (proportional to $J_2$). This uniform $S{=}1$ chain
realizes the Haldane phase: a finite bulk gap and $S=\tfrac12$ edge states
under open boundaries, yielding a near-fourfold ground-state manifold in long
chains that is independent of how the chain is cut.\cite{HaldanePRL1983,HaldanePLA1983,AKLT1987,AKLT1988,PollmannSPT2012}

Finally, periodic electronic-structure calculations place the active manifold deep in the Mott regime, validating a spin-only Heisenberg mapping for \emph{both} realizations. As a dynamical fingerprint we compute $S^{+-}(q,\omega)$: it shows (i) a single triplon branch with bandwidth consistent with $J_1$ in the dimerized $S=\tfrac12$ case and (ii) the Haldane magnon with gap and bandwidth set by the effective inter-monomer antiferromagnetic exchange in the $S=1$ case; open-chain Fourier maps reproduce these ridges with the expected folding, in line with previous studies on different haldane chains.\cite{Harris1973, Bonner1982, White2008, Maeda2007} Detailed explanation regarding the numerical treatment of the spin chains has been provided in the supporting information (SI) section.

\begin{figure*}[t]
  \centering
  \includegraphics[width=\textwidth]{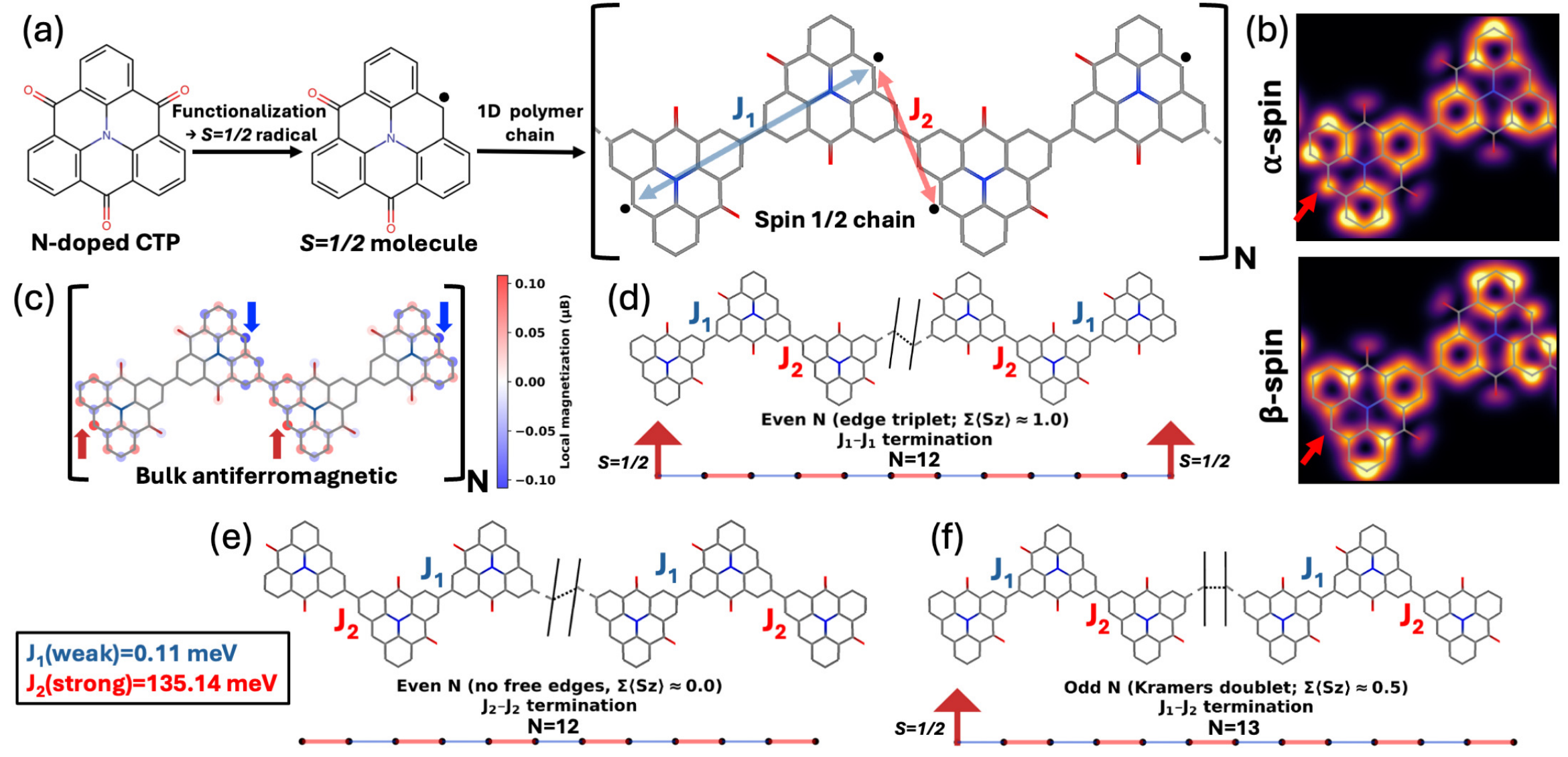}
  \caption{
    Chain chemistry and terminations of the dimerized nanographene
    spin-$1/2$ chain.
    (a) Chemical route from N-doped CTP to a stable $S=1/2$ radical and its
    one-dimensional polymer, whose magnetic backbone is an
    alternating-exchange spin-$1/2$ chain with couplings $J_1$ (blue) and
    $J_2$ (red).
    (b) Simulated spin-polarized STM constant-current maps for the isolated
    radical (top: $\alpha$-spin, bottom: $\beta$-spin), showing complementary
    contrast at the same molecular site (red arrows), consistent with a single
    $S=1/2$ ground state.
    (c) DFT+$U$ local magnetization for a polymer segment, revealing bulk
    antiferromagnetic order with alternating up/down monomer moments.
    (d–f) Schematic open chains with different terminations. $J_1$–$J_1$:
    two $S=1/2$ edge spins forming an edge triplet
    ($\sum_i \langle S_i^z \rangle \approx 1$). $J_2$–$J_2$: intact strong
    dimers and no free edges ($\sum_i \langle S_i^z \rangle \approx 0$).
    Mixed $J_1$–$J_2$: a single unpaired edge spin giving a Kramers doublet
    ($\sum_i \langle S_i^z \rangle \approx 1/2$).
  }
  \label{fig:spinhalf_chain}
\end{figure*}

\section{Results and discussion}

\subsection{Chain chemistry and dimerized $S=1/2$ spin chain}

Figure~\ref{fig:spinhalf_chain} summarizes the chemical design and spin
texture of the nanographene spin-$1/2$ chain. The building block is an
N-doped carbonyl-triphenyl (CTP) core (Fig.~\ref{fig:spinhalf_chain}a, left).
Upon suitable functionalization, one peripheral site is converted into a
stable $S=1/2$ radical (middle), yielding a single unpaired spin largely
confined to each N-doped CTP backbone. Such site-selective radicalization of
the same N-doped CTP motif has been demonstrated for isolated molecules
\cite{WangJACS2022} and for its two-dimensional Kagome lattice on Au(111),
where only a specific carbonyl site hosts the radical.\cite{PawlakACSNano2025}
Here we exploit this selectivity to stitch the radicals into a one-dimensional
chain (right), where neighboring spin centers are coupled along a single
$\pi$-conjugated backbone but connected by two symmetry-inequivalent bridges,
realizing an alternating-exchange Heisenberg chain with weak bond $J_1$ and
strong bond $J_2$.

To visualize the spin texture, we compute spin-resolved LDOS maps within a
DFT+$U$-based Tersoff--Hamann approximation.\cite{Tersoff1998} The simulated
spin-polarized STM images in Fig.~\ref{fig:spinhalf_chain}b show
$\alpha$- and $\beta$-spin contrast that is complementary and localized at the
radical site (red arrows), consistent with a spin-split frontier orbital.
When the radicals are assembled into the polymer, DFT+$U$ calculations reveal
a bulk antiferromagnetic configuration with alternating monomer moments
(Fig.~\ref{fig:spinhalf_chain}c); the color scale highlights small but finite
spin polarization around the radical sites, reflecting the delocalized
$\pi$ character of the spin-$1/2$ state.

Mapping DFT+$U$ total energies onto the Heisenberg model yields
$J_1 \approx 0.11$~meV and $J_2 \approx 135$~meV (see SI section S2), placing the chain extremely
close to the ideal limit of isolated antiferromagnetic dimers. The termination
then controls the edge-state manifold. For an even-$N$ chain with
$J_1$–$J_1$ termination (Fig.~\ref{fig:spinhalf_chain}d), both ends cut weak
bonds and leave two $S=1/2$ edge spins forming an edge triplet with
$\sum_i \langle S_i^z \rangle \approx 1$. For an even-$N$ chain with
$J_2$–$J_2$ termination (Fig.~\ref{fig:spinhalf_chain}e), all strong dimers
remain intact and $\sum_i \langle S_i^z \rangle \approx 0$. An odd-$N$ chain
with mixed $J_1$–$J_2$ termination (Fig.~\ref{fig:spinhalf_chain}f) hosts a
single unpaired edge spin and realizes a Kramers doublet with
$\sum_i \langle S_i^z \rangle \approx 1/2$. These three terminations provide a
direct chemical handle on the edge states of the dimerized spin-$1/2$ chain.

\begin{figure*}[t]
  \centering
  \includegraphics[width=\textwidth]{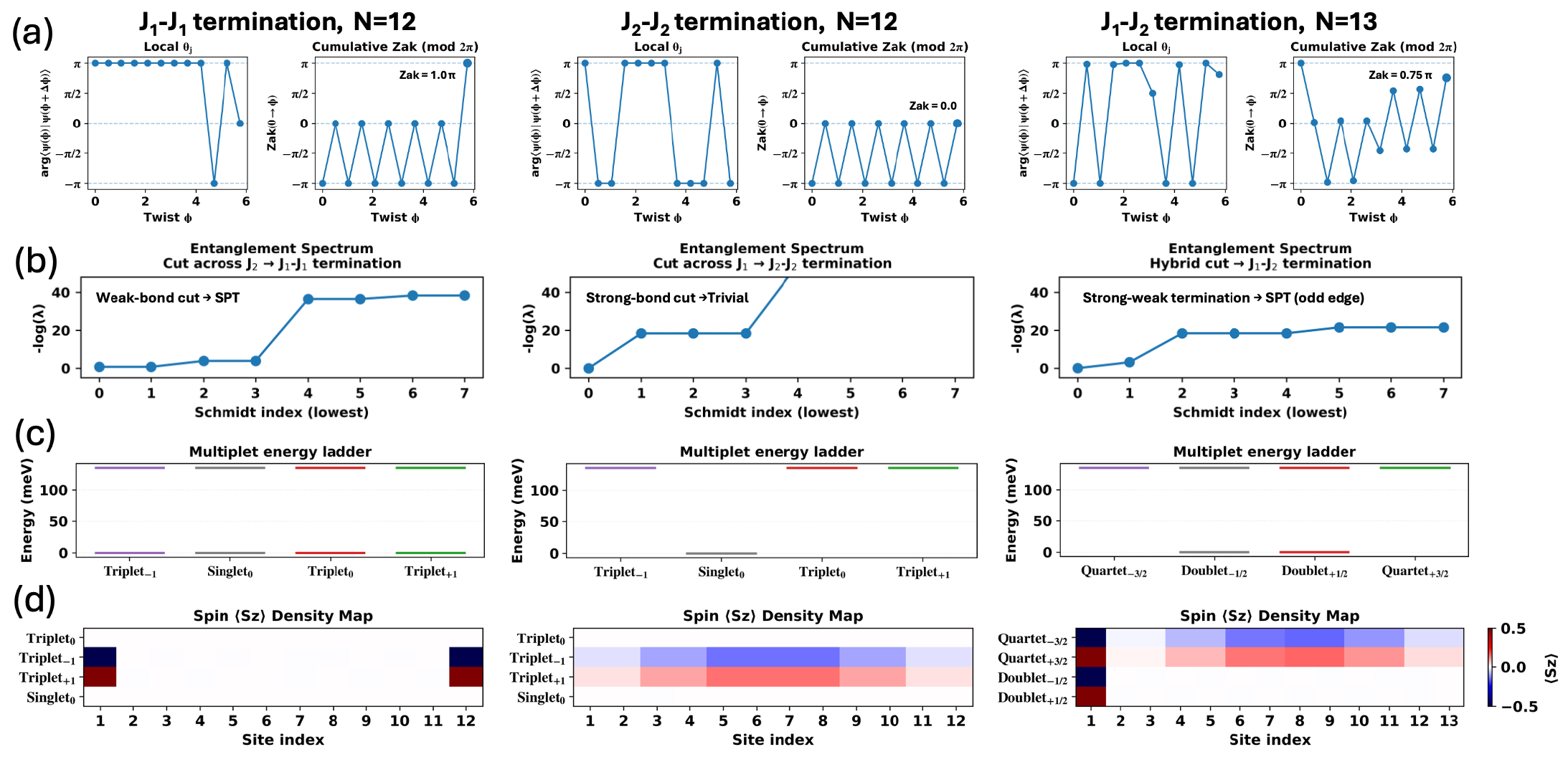}
  \caption{
    Topological diagnostics of the dimerized nanographene spin-$1/2$ chain.
    (a) Many-body Zak phase from twisted boundary conditions. $J_1$–$J_1$
    (even $N$) gives $\pi$ (SPT), $J_2$–$J_2$ (even $N$) gives $0$ (trivial),
    and mixed $J_1$–$J_2$ (odd $N$) yields $\sim 0.75\pi$, consistent with a
    single fractional edge spin-$1/2$.
    (b) Entanglement spectra $-\log\lambda$ versus Schmidt index (with
    $\lambda$ the eigenvalues of the reduced density matrix) for cuts across
    weak ($J_1$) and strong ($J_2$) bonds. Weak-bond cuts in $J_1$–$J_1$ and
    $J_1$–$J_2$ chains show a nearly fourfold-degenerate lowest manifold
    (SPT), whereas a strong-bond cut in $J_2$–$J_2$ chains yields a
    non-degenerate spectrum (trivial).
    (c) Low-energy multiplet ladders from ED: $J_1$–$J_1$ has a triplet ground
    state (two correlated edge spins), $J_2$–$J_2$ a singlet (no free edges),
    and odd-$N$ $J_1$–$J_2$ a Kramers doublet (single edge spin-$1/2$).
    (d) Spin-density maps $\langle S_i^z\rangle$ for the lowest multiplets,
    highlighting two edge-localized $S=1/2$ moments in $J_1$–$J_1$, no edge
    polarization in $J_2$–$J_2$, and a single edge $S=1/2$ in $J_1$–$J_2$.
  }
  \label{fig:spinhalf_topology}
\end{figure*}

\subsection{Topological fingerprints and edge multiplets of the dimerized $S=1/2$ chain}

The different terminations of the nanographene spin-$1/2$ chain lead to
distinct bulk topologies and edge-state manifolds, which we diagnose using
many-body Zak phases, entanglement spectra, and spin-resolved multiplet
structures (Fig.~\ref{fig:spinhalf_topology}). For each geometry we impose a
twist $\phi$ on the periodic bond and follow the ground state
$|\psi_0(\phi)\rangle$ around a loop $0\le\phi<2\pi$. The local Berry angles
between neighboring twists and their cumulative sum
$\mathrm{Zak}(\phi)$ [Fig.~\ref{fig:spinhalf_topology}a] yield a quantized
many-body Zak phase modulo $2\pi$. An even-$N$ chain with $J_1$–$J_1$
termination has Zak phase $\pi$, consistent with an interacting SPT phase
protected by inversion and spin-rotation symmetries, whereas the even-$N$
$J_2$–$J_2$ chain is trivial with Zak phase $0$. For the odd-$N$ chain with
mixed $J_1$–$J_2$ termination, the cumulative phase saturates near
$\sim 0.75\pi$, reflecting finite-size fractionalization of a single edge
spin-$1/2$.

The entanglement spectra (ES) [Fig.~\ref{fig:spinhalf_topology}b], obtained as
$-\log\lambda_n$ from the eigenvalues $\{\lambda_n\}$ of the reduced density
matrix for a bipartition, show the same pattern. Cuts across a weak $J_1$ bond
in $J_1$–$J_1$ and $J_1$–$J_2$ chains exhibit a nearly fourfold-degenerate
lowest manifold, characteristic of an SPT phase, whereas a strong-bond cut in
the $J_2$–$J_2$ chain yields a non-degenerate spectrum typical of a trivial
product state. For the odd-$N$ $J_1$–$J_2$ chain, a hybrid cut produces a
low-lying pair of near-degenerate levels, consistent with a single unscreened
edge spin. The slight deviations from exact fourfold (even-$N$ $J_1$–$J_1$)
or twofold (odd-$N$ $J_1$–$J_2$) degeneracy may arise in ES:
the bipartition of a finite chain creates two inequivalent virtual edges whose
local environments and residual couplings are not strictly symmetric, lifting
the degeneracies slightly while preserving the near-degenerate SPT structure.

To connect these bulk diagnostics to real-space edge physics, we classify the
low-energy eigenstates into SU(2) multiplets and track their energies
[Fig.~\ref{fig:spinhalf_topology}c]. For $J_1$–$J_1$ termination the ground
state is a triplet with degenerate singlet and higher triplet partners, reflecting
two correlated $S=1/2$ edge spins. For $J_2$–$J_2$ termination the ground
state is a singlet separated from triplet excitations, consistent with all
dimers paired into a trivial bulk state. The odd-$N$ $J_1$–$J_2$ chain shows a
Kramers doublet ground state followed by quartet and higher multiplets,
corresponding to a single fractional edge spin-$1/2$.

\begin{figure*}[t]
  \centering
  \includegraphics[width=\textwidth]{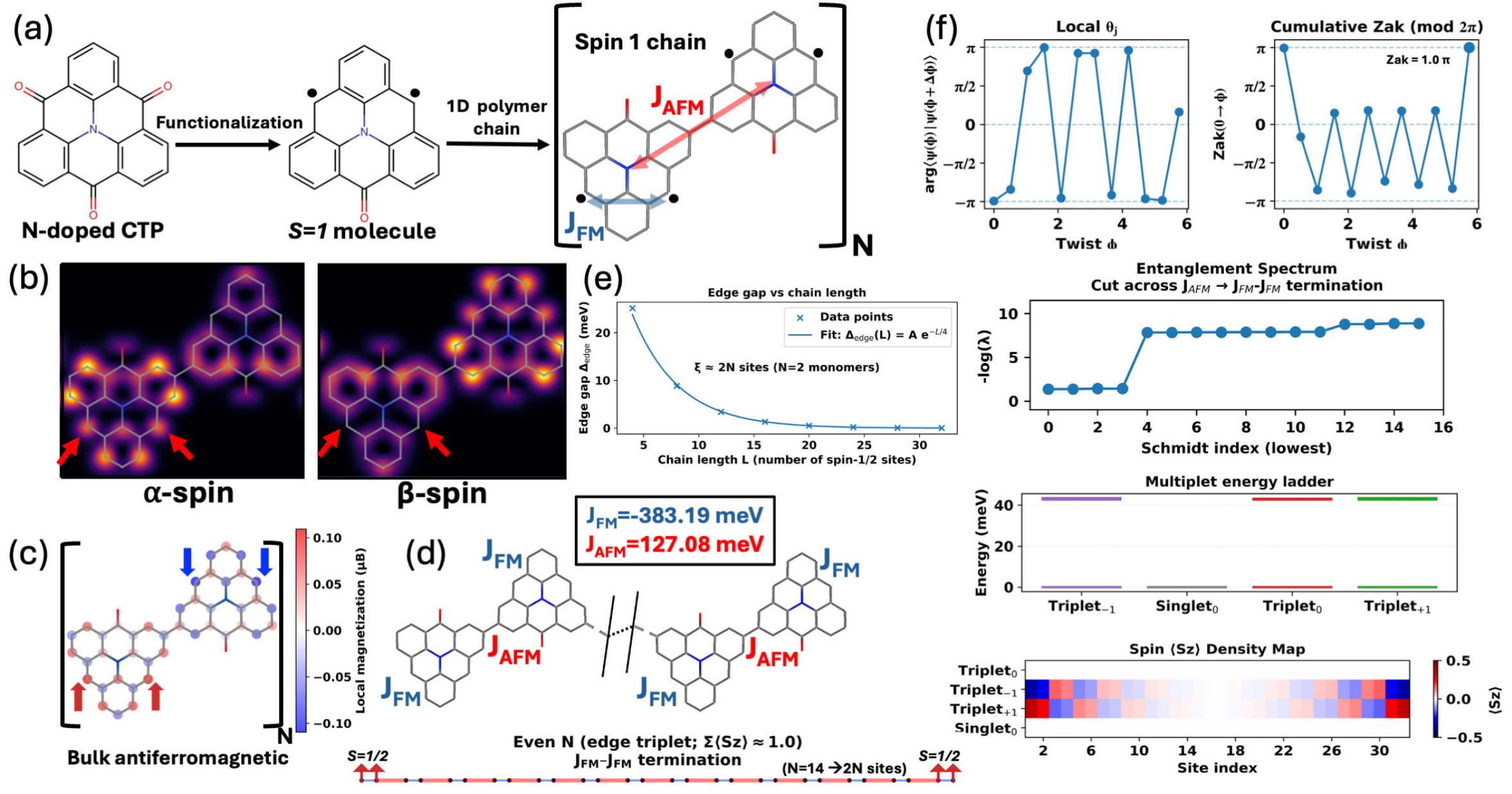}
  \caption{
    Spin-1 Hund chain from N-doped CTP superatoms and its Haldane signatures.
    (a) Chemical route from N-doped CTP to an $S=1$ Hund monomer and its
    one-dimensional polymer, with intra-monomer ferromagnetic
    $J_{\mathrm{FM}}$ and inter-monomer antiferromagnetic
    $J_{\mathrm{AFM}}$ couplings defining the spin-1 chain.
    (b) Simulated spin-resolved STM image of the isolated $S=1$ monomer,
    showing complementary $\alpha$- and $\beta$-spin contrast at the two
    radical sites (red arrows).
    (c) DFT+$U$ local magnetization for a polymer segment, revealing bulk
    antiferromagnetic order with alternating monomer moments.
    (d) Even-$N$ open chain with $J_{\mathrm{FM}}$–$J_{\mathrm{FM}}$
    termination and the corresponding spin-$1/2$ lattice cartoon, exposing two
    edge $S=1/2$ moments that form an edge triplet.
    (e) Edge-triplet splitting $\Delta_{\mathrm{edge}}(L)$ versus chain
    length $L$ with an exponential fit
    $\Delta_{\mathrm{edge}}(L)=A\exp(-L/\xi)$, yielding a short correlation
    length of order two spin-1 monomers.
    (f) DMRG diagnostics: many-body Zak phase quantized to $\pi$, a nearly
    fourfold-degenerate entanglement/multiplet manifold, and spin-density maps
    $\langle S_i^z\rangle$ showing two localized $S=1/2$ edge spins with a
    slight intra-monomer asymmetry.
  }
  \label{fig:spinone_chain}
\end{figure*}

Finally, spin-resolved density maps for the lowest multiplets
[Fig.~\ref{fig:spinhalf_topology}d] make the edge localization explicit. In
the $J_1$–$J_1$ geometry the triplet components show strong, opposite
magnetizations pinned at the two chain ends, with negligible bulk
magnetization. For $J_2$–$J_2$ the spin density is small and nearly uniform,
with no pronounced edge polarization. In the odd-$N$ $J_1$–$J_2$ chain the
ground-state doublet exhibits a localized spin-$1/2$ at a single end of the
chain, realizing the expected fractional edge excitation of an SPT phase with
an odd number of dimers. Together, these diagnostics establish that
weak-bond terminations realize an interacting SPT phase with protected edge
spins, while strong-bond terminations are topologically trivial.

\subsection{Hund-coupled spin-1 chain and Haldane edge states}

We now turn to the spin-1 chain realized by Hund-coupled superatoms built from
the same N-doped CTP motif. As sketched in Fig.~\ref{fig:spinone_chain}a, a
different functionalization stabilizes two radical sites on each molecule,
which are ferromagnetically locked by an intra-molecular exchange
$J_{\mathrm{FM}}<0$ and form an $S=1$ ``Hund monomer''. Stitching these units
into a one-dimensional polymer produces a spin-1 chain in which neighboring
monomers are coupled antiferromagnetically by $J_{\mathrm{AFM}}>0$, i.e.\ a
$J_{\mathrm{FM}}$–$J_{\mathrm{AFM}}$ alternating Heisenberg chain where the
monomer acts as a superatom rather than a rigid atomic $S=1$ ion.

The local spin texture of the $S=1$ monomer is visualized in
Fig.~\ref{fig:spinone_chain}b via computed spin-resolved LDOS maps. The simulated $\alpha$- and $\beta$-spin
STM images show complementary contrast at the two radical sites (red arrows),
consistent with a pair of ferromagnetically aligned $S=1/2$ moments. In the
polymer, DFT+$U$ calculations reveal a bulk antiferromagnetic configuration
with alternating monomer magnetizations and a small but finite spin
polarization extending over the molecular rings
(Fig.~\ref{fig:spinone_chain}c). Mapping collinear DFT+$U$ energies onto the spin model yields a large
intra-monomer ferromagnetic exchange
$J_{\mathrm{FM}}\approx -383$~meV, which locks the two $S=\tfrac12$
spins on each monomer into a Hund-coupled triplet, and an
antiferromagnetic inter-monomer coupling
$J_{\mathrm{AFM}}\approx 127$~meV
(Fig.~\ref{fig:spinone_chain}d), so $|J_{\mathrm{FM}}|\gg J_{\mathrm{AFM}}$
and the system lies deep in the Haldane regime of a spin-1 chain built from
Hund-coupled superatoms.

\begin{figure*}[t]
  \centering
  \includegraphics[width=\textwidth]{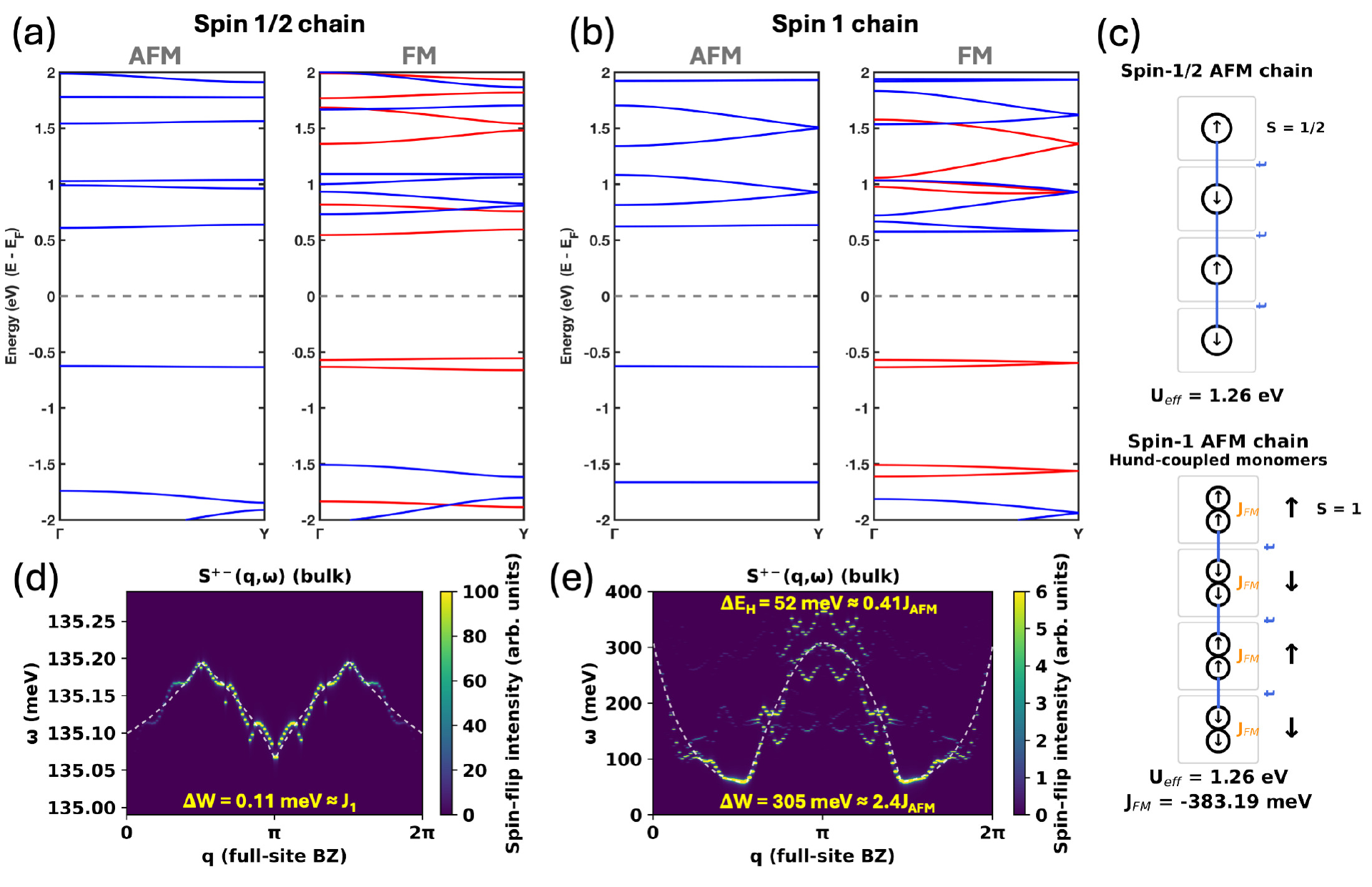}
  \caption{
    (a,b) Spin-resolved DFT+$U$ band structures of the spin-$1/2$ and spin-1
    chains in AFM (left) and FM (right) configurations (blue/red: opposite
    spins; dashed line: $E_F$, bands are spin-degenerate in AFM case). (c) One-orbital Hubbard cartoon: the spin-$1/2$
    chain has one half-filled orbital per monomer with a single hopping $t$
    and on-site interaction $U_{\mathrm{eff}} = 1.26$~eV, while in the spin-1
    chain two ferromagnetically locked orbitals form an $S=1$ site with the
    same $t$ and $U_{\mathrm{eff}}$ and strong intra-monomer exchange
    $J_{\mathrm{FM}}$. (d) Bulk $S^{+-}(q,\omega)$ for the dimerized
    spin-$1/2$ chain, showing a very narrow W-shaped triplon band with
    bandwidth $\Delta W \approx 0.11$~meV $\approx J_1$; the dashed line is
    the one-triplon dispersion of the fitted spin model. (e) $S^{+-}(q,\omega)$
    for the spin-1 Hund chain, displaying an M-shaped triplon branch with
    Haldane gap $\Delta E_H \approx 52$~meV $\approx 0.41 J_{\mathrm{AFM}}$,
    total bandwidth $\Delta W \approx 305$~meV $\approx 2.4 J_{\mathrm{AFM}}$,
    and additional weak branches from internal Hund modes; dashed lines again
    show the one-triplon dispersion.
  }
  \label{fig:bands_triplons}
\end{figure*}

\begin{table*}[t]
  \caption{Exchange parameters, hopping amplitudes, and interaction strengths
  for the spin-$1/2$ and spin-1 chains extracted from DFT+$U$ at 0 and
  300~K.}
  \label{tab:J1J2_tU}
  \centering
  \begin{tabular}{lccccccc}
    \hline
    Species & $T$ (K) & $J_1$ (meV) & $J_2$ (meV) & Characterization
            & $t$ (meV) & $U$ (eV) & $U/t$ \\
    \hline
    Spin $1/2$ & 0   & 0.11      & 135.14 & Dimerized odd Haldane SPT & 12.25 & 1.26 & 103 \\
    \textemdash & 300 & 0.08      & 136.99 & \textemdash               & 11.85 & 1.26 & 106 \\
    Spin 1     & 0   & $-383.19$ & 127.08  & Haldane SPT               & 9.82  & 1.26 & 128 \\
    \textemdash     & 300 & $-376.91$ & 62.70  & \textemdash               & 9.74  & 1.26 & 129 \\
    \hline
  \end{tabular}
\end{table*}

Open chains with an even number of monomers and a
$J_{\mathrm{FM}}$–$J_{\mathrm{FM}}$ termination
(Fig.~\ref{fig:spinone_chain}d) preserve the ferromagnetic dimers at both ends
and cut only antiferromagnetic bonds. This exposes the characteristic
$S=1/2$ edge spins of the Haldane phase: the lowest states form an edge
triplet whose $S^z=+1$ component has
$\sum_i\langle S_i^z\rangle\approx 1$, while the bulk magnetization is
negligible. Because the two $S=1/2$ moments are chemically bound within the
same monomer, the chain can only be terminated between monomers; cutting the
intra-monomer bond would break the molecule. Thus both even- and odd-length
chains always end on intact $S=1$ monomers and are topologically equivalent:
any realistic termination exposes the same pair of fractional $S=1/2$ edge
spins. The finite-size splitting of this edge triplet,
$\Delta_{\mathrm{edge}}(L)$, decays exponentially with chain length $L$,
as shown in Fig.~\ref{fig:spinone_chain}e; a fit
$\Delta_{\mathrm{edge}}(L)=A\exp(-L/\xi)$ yields a correlation length
$\xi\simeq 2N$ sites (about two spin-1 monomers), confirming that relatively
short chains already capture the asymptotic Haldane edge physics.

DMRG calculations \cite{White1992DMRG} on longer chains provide the many-body topological
diagnostics summarized in Fig.~\ref{fig:spinone_chain}f. The many-body Zak
phase obtained from twisted boundary conditions quantizes to $\pi$, signalling
an interacting SPT phase protected by inversion and spin-rotation symmetries,
and the entanglement spectrum shows a nearly fourfold-degenerate lowest
manifold when the cut is placed across the monomers, as expected for a spin-1
Haldane chain. The low-energy multiplet ladder and spin-density maps
(Fig.~\ref{fig:spinone_chain}f, middle and bottom) reveal a triplet ground
state with two localized $S=1/2$ edge spins at opposite ends. Each pair of
neighboring sites in the spin-density map corresponds to the two
ferromagnetically locked $S=1/2$ moments of a single Hund monomer; in the open
chain their individual $\langle S_i^z\rangle$ values differ slightly because
the site facing the inter-monomer antiferromagnetic bond is more strongly
entangled with the neighbor, so finite $|J_{\mathrm{FM}}|/J_{\mathrm{AFM}}$
and boundary effects break the internal symmetry. This slight intra-monomer
imbalance is in fact a key strength of our microscopic description: by keeping
two spin-$1/2$ sites and using the alternating Heisenberg Hamiltonian rather
than a coarse-grained spin-1 model, we resolve the internal structure and
bonding asymmetry of each Hund monomer and closely mirror the real spin
distribution. At the outermost monomers the pair-summed spin density is close
to $\pm 1/2$, identifying the fractional edge $S=1/2$ modes of the Haldane
phase, while interior monomers carry only a small residual net moment.

\subsection{Hubbard mapping, Mott regime, and triplon dynamics}

To connect the \emph{ab initio} electronic structure to the effective spin
models, we first analyze the spin-resolved band structures of the
spin-$1/2$ and spin-1 chains in antiferromagnetic (AFM) and ferromagnetic
(FM) configurations [Fig.~\ref{fig:bands_triplons}a,b]. In both chains the AFM
state is insulating with a large single-particle gap around the Fermi level,
while the FM configuration shows the same bands strongly spin split (red/blue
lines): a single majority-spin band (or pair of bands for the Hund chain) lies
closest to $E_F$, and its dispersion defines the FM bandwidth
$W_{\mathrm{FM}}$ used to extract $t$. While single-particle band structures do not by themselves establish interacting SPT order, 
they provide a compact fingerprint of the real-space connectivity and inversion center that underlie the effective spin models. In the spin-$1/2$ chain, the radicalized site alternates
from one edge of the N-doped CTP monomer to the next, so the effective
$S=1/2$ degrees of freedom are \emph{bond-centered} between monomers; a
translation plus inversion about a bond center generates two inequivalent
exchanges ($J_1,J_2$) and visible band folding and small splittings along
$\Gamma$–Y in Fig.~\ref{fig:bands_triplons}a, the electronic fingerprint of
the frustrated odd-Haldane geometry. By contrast, in the spin-1 chain the
Hund-coupled superatoms are \emph{site-centered} and all AFM links between
monomers are symmetry-equivalent, leading to more regular, nearly symmetric
AFM and FM band ladders [Fig.~\ref{fig:bands_triplons}b] consistent with a
uniform site-centered spin-1 Haldane chain.

Following the standard bandwidth scaling of one-band Hubbard mappings,%
\cite{Xiang2013}
we extract the nearest-neighbor hopping $t$ from the FM bandwidth of a single
spin channel,
\begin{equation}
  W_{\mathrm{FM}} = 2 z\,|t|
  \;\Rightarrow\;
  |t| = \frac{W_{\mathrm{FM}}}{2 z},
\end{equation}
$z=2\ \text{(1D)} \Rightarrow |t| = \frac{W_{\mathrm{FM}}}{4}$.
and estimate the effective on-site repulsion from the AFM spectra via
\begin{equation}
  U_{\mathrm{eff}} \approx E_g^{\mathrm{AFM}} + W_{\mathrm{AFM}} .
\end{equation}
These relations give the usual correlation measure $U_{\mathrm{eff}}/t$ used
to locate the Mott regime in $U/t$ space ($U/t \ge 4.3$).\cite{Meng2010, ANINDYA2022} Applying this
procedure to the DFT+$U$ bands in Fig.~\ref{fig:bands_triplons}a,b yields
$t\approx 12.25$~meV for the spin-$1/2$ chain and $t\approx 9.82$~meV for the
spin-1 chain at 0~K, with an essentially identical
$U_{\mathrm{eff}}\approx 1.26$~eV for both. The resulting ratios
$U_{\mathrm{eff}}/t \sim 100$–130 place both chains deep in the Mott regime,
fully justifying the spin-only Heisenberg description. The corresponding
parameters at 0 and 300~K are summarized in Table~\ref{tab:J1J2_tU}.

Panel~\ref{fig:bands_triplons}c summarizes the underlying one-band Hubbard
picture. For the spin-$1/2$ chain, each radical monomer is represented by a
single half-filled orbital with on-site interaction
$U_{\mathrm{eff}}=1.26$~eV and a single nearest-neighbor hopping $t$ between
monomers. In the spin-1 chain, two such orbitals on each monomer are
ferromagnetically locked into an effective $S=1$ superatom, but charge
fluctuations between neighboring monomers are still governed by the same
scalar hopping $t$ and on-site $U_{\mathrm{eff}}$. Thus both the dimerized
spin-$1/2$ chain and the Hund-coupled spin-1 chain descend from the same
fermionic backbone with one effective inter-monomer hopping and a common
$U_{\mathrm{eff}}$, differing only in local spin multiplicity and the
resulting exchange hierarchy.

To assess thermal robustness, we evolve each chain at 300~K in LAMMPS using
SevenNet-MF-ompa machine-learning (ML) interatomic potentials (see SI section S3 and S4),%
\cite{Kim2025,Park2024}
fine-tuned with DFT data for the same molecular family.\cite{Anindya2025}
Representative configurations from these molecular-dynamics trajectories are
quenched and re-evaluated with the same DFT+$U$ band-structure and
exchange-mapping procedure. The finite-temperature parameters
(Table~\ref{tab:J1J2_tU}) show only modest renormalization: $t$ is reduced by
$\sim 3$–5\% (from 12.25 to 11.85~meV in the spin-$1/2$ chain and from 9.82 to
9.74~meV in the spin-1 chain), while $J_2$ changes by less than 2~meV and
$U_{\mathrm{eff}}$ remains essentially unchanged. Thus, the chains remain deep in the Mott regime, 
and thermal distortions at 300 K do not drive the system out of the exchange hierarchy 
and correlation regime that supports the odd-Haldane and Haldane SPT ground states.

Finally, we connect this Hubbard-scale picture to the many-body spin dynamics
via the bulk spin-flip structure factor $S^{+-}(q,\omega)$
[Fig.~\ref{fig:bands_triplons}d,e]. For the dimerized spin-$1/2$ chain,
$S^{+-}(q,\omega)$ shows a very narrow triplon band with total bandwidth
$\Delta W \approx 0.11$~meV, matching the weak inter-dimer coupling
$J_1$,\cite{Bonner1982} and displaying a characteristic ``W''-shaped dispersion: the band
maximum lies near $q\!\approx\!\pi/2$ (and $3\pi/2$) and the minima near
$q=0,\pi,2\pi$, with most spectral weight between $0$ and $\pi$. This pattern
reflects the bond-centered odd-Haldane chain: the elementary triplon lives on
a strong $J_2$ dimer, so the effective dimer lattice has twice the spacing of
the site lattice and its AFM wave vector maps to $q\!\approx\!\pi/2$ in the
full-site Brillouin zone, while form-factor cancellations suppress weight at
$q=0,\pi$. In contrast, the Hund-coupled spin-1 chain displays a broad triplon
branch with an ``M''-shaped dispersion: the band minimum lies near
$q\!\approx\!\pi/2$, the maxima occur near $q\!\approx\!0$ and $\pi$, the
Haldane gap at the minimum is
$\Delta E_H \approx 52$~meV $\simeq 0.41 J_{\mathrm{AFM}}$, and the total
bandwidth $\Delta W \approx 305$~meV is of order
$2.4 J_{\mathrm{AFM}}$.\cite{White2008, Maeda2007} Additional, weaker branches at higher energies reflect secondary
triplon modes associated with the internal Hund-coupled structure, underscoring
that the chain is not a perfectly rigid spin-1 model but a molecular superatom
realization. The dashed ridges overlaying both intensity maps are the
one-triplon dispersions obtained from the ED spin models with
$J_{\mathrm{FM}}$ and $J_{\mathrm{AFM}}$ fitted to DFT+$U$, demonstrating
quantitative agreement between the microscopic Hubbard-scale description and
the emergent triplon dynamics.

\section*{Conclusions}

We have shown that a single N-doped carbonyl-triphenyl platform can host two
distinct interacting SPT phases in one dimension: a bond-centered,
dimerized spin-$1/2$ odd-Haldane chain and a site-centered spin-1 Haldane
chain built from Hund-coupled superatoms. Spin-polarized DFT+$U$ combined
with exchange mapping, ED, DMRG, and $S^{+-}(q,\omega)$ demonstrates that
both systems lie deep in the Mott regime, exhibit fractional $S=1/2$ edge
modes, and display characteristic W- and M-shaped triplon bands in
quantitative agreement with fitted Heisenberg models. ML-accelerated
finite-temperature calculations further show that $J_1$, $J_2$, $t$, and
$U_{\mathrm{eff}}$ are only weakly renormalized at 300~K, indicating that the microscopic exchange hierarchy and correlation regime 
underpinning the SPT ground states is stable against thermal distortions up to room temperature and broadly
promising as chemically programmable platforms for edge-spin qubits,
topological spin transport, and quantum simulation of interacting SPT phases.

\section{Methods}

The effective spin models are isotropic Heisenberg chains and are therefore
SU(2) symmetric at zero field. \cite{AuerbachBook1994, Gandon2025} In practice we work in fixed
$S^z_\mathrm{tot}$ sectors and verify a posteriori that the low-lying states
form nearly degenerate SU(2) multiplets by monitoring
$\langle \hat{\mathbf{S}}_\mathrm{tot}^2\rangle$ for the corresponding
eigenstates.

For the strongly dimerized spin-$1/2$ alternating-exchange chain relevant to
the nanographene edge segments, the intra-monomer coupling is effectively
vanishing, $J_1 \approx 0$, so the system is close to perfectly isolated
antiferromagnetic dimers. The correlation length is then only a few lattice
sites, and a chain of $N=12$ monomers already eliminates noticeable
finite-size effects: exact diagonalization (ED) on this size yields converged
gaps and one-triplon dispersions indistinguishable, within numerical
accuracy, from longer chains. In contrast, for the effective spin-1 chain
realized by Hund-coupled superatoms the intra-monomer ferromagnetic exchange
is large but finite, $|J_1| \gg J_2$, so the monomers are not ideal rigid
$S=1$ atoms and the Haldane edge states extend over several units. To capture
this we study chains of $N=32$ spin-$1/2$ sites (16 Hund monomers), treated
with density-matrix renormalization group (DMRG) using the TenPy library and
exploiting only the U(1) symmetry associated with $S^z_\mathrm{tot}$. ED on
short chains and DMRG on longer ones provide consistent edge gaps, bulk gaps,
and triplon dispersions.

The dynamical spin structure factor $S^{+-}(q,\omega)$ is computed on
periodic chains via a Lanczos continued-fraction evaluation of the
zero-temperature correlation function, starting from the exact ground state
and acting with
$\hat{S}_q^+ = N^{-1/2} \sum_j e^{-iq j} \left(\hat{S}_j^x + i \hat{S}_j^y\right)$.
The resulting intensity maps capture the bulk triplon bands and their dependence on the microscopic couplings, and are the theoretical counterparts of spin-flip features in STM/STS and magnon branches in inelastic neutron scattering. \cite{HirjibehedinScience2006, Lovesey1984}

The exchange parameters $J_1$ and $J_2$ are obtained by mapping collinear
spin-polarized DFT+$U$ total energies. All energies entering this mapping are
computed with element-resolved Hubbard corrections on the C, N, and O sites of
the CTP backbone,\cite{PawlakACSNano2025} determined from first-principles
linear-response calculations in which small on-site potential shifts are
applied and the induced changes in orbital occupations are fitted.\cite{Cococcioni2005}
This treatment is particularly important for the localized carbonyl O $2p$
states, which are sensitive to self-interaction errors and control the local
moments and exchange splittings.

For the spin-$1/2$ alternating-exchange chain, each monomer hosts a single
localized $S=1/2$ moment and the low-energy physics is described by two
antiferromagnetic couplings $J_1,J_2>0$ along the chain. We construct a
four-site supercell and evaluate the DFT energies of several collinear
configurations (\,$\uparrow\uparrow\vert\uparrow\uparrow$,
$\uparrow\downarrow\vert\uparrow\downarrow$,
$\uparrow\uparrow\vert\downarrow\downarrow$, etc.). Expressing these energies
in terms of $\langle \mathbf{S}_i\!\cdot\!\mathbf{S}_j\rangle=\pm\tfrac14$ for
each bond and solving the resulting linear system yields $J_1$ (the weaker
bond) and $J_2$ (the stronger bond); in our nanographene chain we find
$J_1\simeq 0$ and $J_2>0$, placing the system very close to the ideal
dimer limit. For the effective spin-1 chain realized by Hund-coupled
superatoms, each monomer hosts two ferromagnetically locked $S=1/2$ moments.
The intra-monomer splitting between FM and AFM alignment defines a large
ferromagnetic Hund exchange $J_{\mathrm{FM}}<0$, while the energy difference
between parallel and antiparallel alignment of neighboring monomers in a
two-monomer cell defines an antiferromagnetic inter-monomer exchange
$J_{\mathrm{AFM}}>0$. In the spin-1 description we therefore identify
\begin{equation}
  J_1 = J_{\mathrm{FM}}, \qquad
  J_2 = J_{\mathrm{AFM}} \equiv J_{\mathrm{eff}},
\end{equation}
where $J_{\mathrm{eff}}$ is the effective antiferromagnetic coupling between
neighboring Hund-coupled $S=1$ superatoms that sets the Haldane gap and
triplon bandwidth. Further details of the methods are described in the SI section.

\bibliography{acs-latex-template}        

\section*{Acknowledgements}

This work was supported by the Natural Sciences and Engineering Research Council of Canada (NSERC) and Fonds de recherche du Québec – Nature et technologies (FRQNT) and would have not been possible without the computational resources provided by Calcul Québec and the Digital Research Alliance of Canada (The Alliance).

\section*{Supporting information}

The Supporting Information is available free of charge at ......

Model symmetries and numerical methods for the spin-chain calculations (Section S1); DFT+$U$ setup and extraction of exchange parameters for the spin-$\tfrac{1}{2}$ and Hund-coupled spin-1 chains (Section S2); finite-temperature molecular dynamics and construction of time-averaged structures (Section S3); and 300~K structural stability diagnostics (Section S4; Supporting Figure S1).



%
%
%
%
%
\end{multicols}

\end{document}


\maketitle

%


\section*{S1. Model symmetries and numerical methods}

\subsection*{S1.1 SU(2) symmetry and sector structure}

The spin model used in the main text is the isotropic, nearest-neighbor
Heisenberg chain with alternating couplings [Eq.~(1) in main text]. At zero magnetic
field this Hamiltonian is fully SU(2) symmetric: it commutes with the
components of the total spin
\(
\hat{\mathbf{S}}_{\mathrm{tot}}
= \sum_i \hat{\mathbf{S}}_i
\)
and with
\(
\hat{\mathbf{S}}_{\mathrm{tot}}^2
\),
so eigenstates can in principle be labeled by $(S,S^z)$.

In both our exact-diagonalization (ED) and DMRG calculations we exploit
the U(1) subgroup generated by $\hat{S}^z_{\mathrm{tot}}$. The Hilbert
space is decomposed into blocks with fixed $S^z_{\mathrm{tot}}$, and the
many-body problem is solved independently in each block. The full SU(2)
structure is then used a posteriori: for low-lying eigenstates
$|\psi_n\rangle$ we compute
\(
\langle \hat{\mathbf{S}}_{\mathrm{tot}}^2\rangle_n
\)
and check that they cluster near $S(S+1)$ with $S=0,1,\ldots$, confirming
the expected singlet, triplet, etc.\ multiplets and their near
degeneracies. This is particularly important for identifying the Haldane
phase, where the ground state in the bulk is a singlet but the open
chain exhibits emergent $S=1/2$ edge degrees of freedom.

Whenever we probe a finite magnetic field $B$ along $z$ we add the Zeeman
term $-g\mu_B B \sum_i \hat{S}_i^z$ on top of Eq.~(1). This breaks SU(2)
down to U(1), but $\hat{S}^z_{\mathrm{tot}}$ remains conserved so the
block structure is unchanged. The field simply splits the SU(2)
multiplets in the expected way and can be used to identify level
crossings and saturation fields.

\subsection*{S1.2 ED for the strongly dimerized spin-$\tfrac{1}{2}$ chain}

For the “odd-Haldane” spin-$1/2$ realization relevant to the
nanographene spin chains, the mapping from DFT yields a very strongly
dimerized limit: the weak coupling $J_1$ is effectively
vanishing ($J_1 \approx 0$) while the strong coupling $J_2>0$
sets the main energy scale. Physically, this corresponds to almost
perfect dimers weakly coupled into a chain. The correlation length is
then very short, of order a couple of sites, so finite-size effects decay
rapidly with chain length.

In this regime we treat the system entirely with exact diagonalization:

\begin{itemize}
  \item We work with an open chain of
        \(
        N = 12
        \)
        spin-$1/2$ sites. Here one spin-$1/2$ site corresponds to one
        nanographene monomer, so $N=12$ corresponds to a chain of 12
        monomers.

  \item The Hilbert space dimension is $2^{12} = 4096$, which allows us
        to construct the full Hamiltonian of Eq.~(1) in sparse-matrix
        form in each $S^z_{\mathrm{tot}}$ sector and diagonalize it using
        Lanczos (or Arnoldi) routines. We typically target the lowest
        few eigenstates in each block.

  \item We use open boundary conditions with the same alternating bond
        pattern as in the real chain. The edges cut through a weak bond,
        leaving the strong (dimer) bonds just inside the chain and
        exposing fractionalized $S=1/2$ edge degrees of freedom.

  \item The ground state is obtained in the lowest-$|S^z_{\mathrm{tot}}|$
        sector ($S^z_{\mathrm{tot}}=0$ at $B=0$). Excited states in
        $S^z_{\mathrm{tot}} = \pm 1$ are used to define bulk and edge
        triplet gaps.
\end{itemize}

Because numerical diagonalization in $S^z_{\mathrm{tot}}=0$ does not
by itself enforce any constraint on the spatial distribution of
$\langle \hat{S}_i^z\rangle$, we additionally symmetrize the ground
state with respect to a global $\pi$ rotation around the $x$ axis,
\(
\hat{F} = \bigotimes_i \hat{\sigma}_i^x
\)
(where $\hat{\sigma}_i^x$ is the Pauli matrix acting on site $i$).
Given a numerically obtained $S^z_{\mathrm{tot}}=0$ eigenvector
$|\psi\rangle$, we form
\(
|\psi_{\pm}\rangle \propto (|\psi\rangle \pm \hat{F}|\psi\rangle)
\)
and pick the lower-energy parity sector. This projection enforces
$\langle \hat{S}_i^z\rangle \equiv 0$ in the bulk Haldane region of the
spin-$1/2$ chain and makes the emergence of edge magnetization near the
boundaries much cleaner.

From these ED calculations on $N=12$ we extract:

\begin{enumerate}
  \item The bulk triplon gap and dispersion bandwidth (via the lowest
        $S^z_{\mathrm{tot}}=\pm1$ excitations on a ring; see below).
  \item The open-chain edge gap (difference between the lowest singlet
        and lowest triplet for the same $S^z_{\mathrm{tot}}$).
  \item Real-space spin densities $\langle \hat{S}_i^z\rangle$, which
        show the fractionalized edge spins and their decay into the
        bulk.
  \item The spectrum of low-lying states as a function of
        $S^z_{\mathrm{tot}}$, which we organize into multiplets using
        $\langle \hat{\mathbf{S}}_{\mathrm{tot}}^2\rangle$.
\end{enumerate}

We verified that increasing the chain length to $N=14$ and $N=16$ in the
same dimerized regime does not modify the gaps significantly. Thus, $N=12$ monomers already behaves as an effectively infinite chain for the
spin-$1/2$ case.

\subsection*{S1.3 Periodic chain, triplon dispersion, and $S^{+-}(q,\omega)$}

To access momentum-resolved information we also consider the periodic
version of the chain. On a ring we implement boundary conditions and
“twists” by modifying the bond between the last and first sites in
Eq.~(1). A twist angle $\phi$ is introduced on the wrap bond in the
transverse exchange part, so that the many-body ground state and lowest
$S^z_{\mathrm{tot}}=\pm1$ excitations can be followed as a function of
$\phi$. We use the boundary twist $\phi$ as a continuous proxy for the crystal momentum in the reduced (two-site) Brillouin zone, folding $k$ into $0 \le k \le \pi$. Here $k$ labels the dispersion extracted from twisted boundary conditions, while $q \in [0,2\pi)$ is the wavevector used in $S^{+-}(q,\omega)$ on the site lattice.

For each twist angle we compute:

\begin{itemize}
  \item The ground-state energy $E_0(\phi)$ in the sector with smallest
        $|S^z_{\mathrm{tot}}|$ (typically $S^z_{\mathrm{tot}}=0$).
  \item The lowest excitation $E_1^{(+)}(\phi)$ in the sector
        $S^z_{\mathrm{tot}}=+1$.
\end{itemize}

The one-triplon dispersion is then defined as
\begin{equation}
E_{\mathrm{bulk}}(k) = E_1^{(+)}(\phi) - E_0(\phi), \qquad k \equiv \phi.
\end{equation}
This is what we compare to the dominant ridge in the dynamical structure
factor $S^{+-}(q,\omega)$ and to analytical expectations based on the strong-dimer (triplon) description of the alternating-exchange chain
and effective Haldane chain.

For the dynamical spin structure factor itself, we evaluate
\begin{equation}
S^{+-}(q,\omega)
=
\frac{1}{\pi} \sum_n
\left|\langle n | \hat{S}_q^+ | 0 \rangle\right|^2
\frac{\eta}{[\omega - (E_n - E_0)]^2 + \eta^2},
\end{equation}
where $|0\rangle$ is the ground state at $\phi=0$,
$\hat{S}_q^+ = N^{-1/2} \sum_j e^{-iq j}(\hat{S}_j^x + i\hat{S}_j^y)$
is the Fourier-transformed spin-raising operator, and $\eta$ is a small
broadening. Numerically, this is implemented with a Lanczos
continued-fraction method starting from the initial state
$|\phi_q\rangle = \hat{S}_q^+ |0\rangle$. The Krylov subspace generated
by successive actions of the Hamiltonian is tridiagonalized, and the
continued fraction for the local Green’s function is evaluated on a grid
of $\omega$. In practice we use on the order of $10^2$–$10^3$ Lanczos
iterations, a uniform grid of $q$ values in $[0,2\pi)$, and a broadening
$\eta$ of order $0.5$–$1$~meV. The resulting intensity maps are lightly
smoothed along the energy axis and used to extract the ridge that
defines the triplon band in the main text.

\subsection*{S1.4 DMRG for the Hund-coupled spin-1 chain}

The “Haldane chain” realized in the nanographene $S=1$ platform is not a
chain of ideal spin-1 atoms, but a chain of Hund-coupled superatoms:
each monomer hosts two spin-$1/2$ sites, coupled ferromagnetically by a
large $J_1<0$, and neighboring monomers are coupled antiferromagnetically
by $J_2>0$ with $|J_1| \gg J_2$. In the limit $|J_1|\to\infty$ each
monomer would form an exact $S=1$ unit, but at realistic $J_1$ the
local moment is slightly “soft” and the correlation length is enhanced
relative to the ideal spin-1 chain.

To capture this realistic spin-1 Haldane physics we need substantially
larger systems than are accessible to ED. We therefore use DMRG \cite{White1992DMRG} as
implemented in the TenPy library \cite{Hauschild2018Tenpy}, working with chains of
\[
N_{\text{sites}} = 32
\]
spin-$1/2$ sites grouped into 16 monomers (two sites per monomer). In
the effective spin-1 description this corresponds to a chain of
$N_{S=1} = 16$ sites.

The main features of the DMRG setup are:

\begin{itemize}
  \item We retain the U(1) symmetry associated with conservation of
        $S^z_{\mathrm{tot}}$. The MPS tensors therefore carry a U(1)
        “charge” and the Hilbert space is block-sparse, which greatly
        improves efficiency. Non-Abelian SU(2) is not enforced at the
        tensor-network level, but we again verify that the low-energy
        spectrum organizes into approximate SU(2) multiplets via
        $\langle \hat{\mathbf{S}}_{\mathrm{tot}}^2\rangle$.

  \item We use open boundary conditions and the same alternating bond
        pattern as in the real Hund chain: strong ferromagnetic $J_1$
        bonds within each monomer and antiferromagnetic $J_2$ bonds
        between monomers. This produces the expected Haldane edge states,
        whose spatial profiles are extended across several physical sites
        near each end.

  \item We perform finite-system DMRG sweeps, increasing the maximum bond
        dimension up to $\chi \sim 10^3$ and converging the truncation
        error to $\epsilon_{\mathrm{trunc}} \leq 10^{-8}$ for the
        ground state. Convergence is checked by monitoring the energy,
        entanglement entropy and local observables as functions of sweep
        number and $\chi$.
\end{itemize}

From the converged MPS we obtain:

\begin{enumerate}
  \item The bulk Haldane gap, by targeting the lowest state in the
        lowest-$|S^z_{\mathrm{tot}}|$ sector and the lowest state in
        $S^z_{\mathrm{tot}}=\pm 1$, taking their energy difference and
        extrapolating in the centre of the chain.

  \item The edge magnetization profile
        $\langle \hat{S}_i^z\rangle$, which shows the emergent
        fractionalized edge spins and their exponential decay into the
        bulk. Fitting this decay yields a correlation length $\xi$ that
        is a couple monomers long.

  \item The entanglement spectrum across a cut in the middle of the
        chain. The even-fold degeneracy of the Schmidt values provides a
        robust diagnostic of the Haldane SPT phase and is consistent
        with the edge and bulk gap analysis.
\end{enumerate}

Test runs at $N_{\text{sites}} = 40$ (20 monomers) show that bulk
quantities (gap, entanglement spectrum, central $\langle \hat{S}_i^z\rangle$)
are indistinguishable from the $N_{\text{sites}} = 32$ results within
numerical accuracy. We therefore use $N_{\text{sites}}=32$ as our
standard system size for the Hund-coupled spin-1 chain.

\subsection*{S1.5 Many-body Zak phase and string order parameter}

For completeness. we have
implemented standard SPT diagnostics such as the many-body Zak (Berry)
phase and the nonlocal string order parameter. We determined Zak-phase based on inserting a global
twist $\phi$ on the ring and following the many-body ground state
$|\psi_0(\phi)\rangle$ along a closed loop $0 \le \phi < 2\pi$. The many-body
Zak (Berry) phase is obtained from the discretized Berry connection
\begin{equation}
  \gamma_\mathrm{Zak}
  =
  - \mathrm{Im}
  \sum_{j}
  \ln \frac{\langle\psi_0(\phi_j) | \psi_0(\phi_{j+1})\rangle}
           {|\langle\psi_0(\phi_j) | \psi_0(\phi_{j+1})\rangle|},
\end{equation}
which is quantized to $0$ or $\pi$ (mod $2\pi$) in the presence of the
protecting symmetries and cleanly distinguishes the topologically trivial and
Haldane phases. In addition to the results shown in main manuscript, we have also computed the nonlocal string order parameter which is evaluated directly
from ED and DMRG, and shows the expected finite value in the SPT regime and
its collapse upon tuning to the trivial phase. These quantities are fully consistent with the
edge/bulk gap diagnostics above and provide an independent many-body
confirmation of the interacting SPT phase.

\subsection*{S1.6 Consistency between ED and DMRG}

The ED and DMRG calculations address complementary regimes of the same
effective spin model. For the strongly dimerized spin-$\tfrac{1}{2}$ chain
($J_1 \approx 0$) ED on $N=12$ monomers is essentially exact and provides
unbiased access to the full low-energy spectrum, edge gaps, triplon
bandwidth, and real-space spin profiles. In this regime the correlation
length is so short that increasing the system size to $N=14$ or $N=16$ has
a negligible effect, indicating that ED already represents the
thermodynamic limit.

For the Hund-coupled spin-1 chain, an exponential fit of the edge gap
vs chain length yields a correlation length $\xi \approx 4$ spin-$\tfrac{1}{2}$
sites $\approx 2$ Hund monomers; the edge spins thus decay over only a
few monomers. To obtain a clean bulk region and essentially decoupled
edges we therefore simulate chains of 32 spin-$\tfrac{1}{2}$ sites
(16 monomers), which is beyond the reach of ED and is treated with
DMRG (TenPy) exploiting $\mathrm{U}(1)$ symmetry. Here DMRG on
$N_{\mathrm{sites}}=32$ (16 monomers) provides accurate bulk gaps,
entanglement spectra, and edge profiles. In parameter regimes where both
methods are feasible (e.g.\ reduced $|J_1|$ and shorter chains), we find
that ED and DMRG yield consistent gaps and local observables within
numerical accuracy. This cross-check supports the use of ED for the
spin-$\tfrac{1}{2}$ chain and DMRG for the Hund-coupled spin-1 chain as
reliable and mutually consistent descriptions of the interacting SPT
phases discussed in the main text.

\section*{S2. DFT+$U$ calculations and extraction of exchange parameters}

\subsection*{S2.1 DFT+$U$ setup}

All electronic-structure calculations are performed within spin-polarized
DFT+$U$ using the PBE generalized-gradient approximation as implemented in
the \textsc{Quantum ESPRESSO} package.\cite{Giannozzi_2009, Giannozzi_2017} Such DFT+$U$ approaches have been validated previously for related systems.\cite{ANINDYA2022, PawlakACSNano2025} The interaction between
valence electrons and ionic cores is described by projector-augmented
wave (PAW) datasets.\cite{BlochPAW1994}

The unit cells of the CTP-based chains contain
64 atoms (including hydrogen terminations). All structures are fully relaxed until the
total-energy change is smaller than $10^{-7}$~Ry and the residual forces
on each atom are below $10^{-6}$~Ry/\AA. Structural optimizations are
carried out on a $\Gamma$-centered $8\times 1\times 1$ $k$-point mesh,
with a plane-wave kinetic-energy cutoff of 50~Ry. A vacuum spacing of
20~\AA\ is used along the out-of-plane ($z$) direction to avoid spurious
interactions between periodic images. All spin-polarized calculations
assume collinear magnetism with the spin quantization axis aligned along
$z$ for simplicity.

For accurate band structures and density of states we use a denser
$30\times 1\times 1$ $k$-point mesh on the relaxed geometries. 

\subsection*{S2.2 Linear-response Hubbard $U$ on C, N, and O sites}

The Hubbard corrections in our DFT+$U$ calculations are applied to the
$2p$ shells of the C, N, and O atoms that belong to the $\pi$-conjugated
backbone and carbonyl groups of the CTP framework. The corresponding
on-site interaction parameters $U_\alpha$ (with $\alpha=\mathrm{C},\mathrm{N},
\mathrm{O}$) are obtained from first-principles linear-response theory
following the Cococcioni and de~Gironcoli scheme.\cite{Cococcioni2005} We use
the implementation provided in the \textsc{Quantum ESPRESSO} distribution.

In this approach, one introduces a small on-site potential shift
$\delta V_\alpha$ on the localized orbital manifold of a given atomic
species (e.g.\ the $2p$ subspace of a particular carbon site) and computes
the induced change in the corresponding occupation numbers $n_\alpha$.
From these we construct the bare and screened response matrices,
\[
\chi^{0}_{\alpha\beta} = \frac{\partial n_\alpha}{\partial V_\beta}
\Big|_{\mathrm{KS}}, \qquad
\chi_{\alpha\beta} = \frac{\partial n_\alpha}{\partial V_\beta}
\Big|_{\mathrm{SCF}},
\]
where $\alpha,\beta$ run over all symmetry-inequivalent Hubbard sites in
the unit cell. The effective Hubbard matrix is then obtained as
\[
U_{\alpha\beta}
=
\left( \chi^{0^{-1}} - \chi^{-1} \right)_{\alpha\beta},
\]
and the on-site $U_\alpha$ entering the DFT+$U$ functional is taken as
the diagonal element $U_{\alpha\alpha}$, averaged over symmetry-equivalent
atoms. In practice we find that the off-diagonal elements are small and
that $U_\alpha$ is nearly site-independent within each chemical species,
so we use a single $U_{\mathrm{C}}$, $U_{\mathrm{N}}$, and $U_{\mathrm{O}}$
for all equivalent backbone and carbonyl sites.\cite{PawlakACSNano2025}

This procedure is particularly important for the carbonyl O $2p$ states,
whose self-interaction errors strongly influence the local moment on the
adjacent CTP backbone and the magnitude of the exchange splitting. With
the linear-response $U$ values, the DFT+$U$ band structures correctly
recover the open-shell character and the magnitude of the experimental
spin splittings reported for related systems.\cite{PawlakACSNano2025}

\subsection*{S2.3 Mapping collinear DFT+$U$ energies to $J_1,J_2$:
spin-$\tfrac{1}{2}$ alternating-exchange chain}

For the spin-$1/2$ alternating-exchange chain the low-energy physics is
captured by two antiferromagnetic couplings $J_1,J_2>0$ along the polymer
backbone. Each monomer hosts a single localized spin-$1/2$ moment, and
the effective one-dimensional Heisenberg model [Eq.~(1)] contains two
distinct nearest-neighbor bonds per unit cell.

To extract $J_1$ and $J_2$ we construct a four-site supercell containing
two unit cells of the spin chain. We then compute DFT+$U$ total energies
for a set of collinear spin configurations, chosen such that each bond is
sampled in both ferromagnetic (FM) and antiferromagnetic (AFM)
alignments. A convenient choice is:
\begin{align*}
  \mathrm{FM} &: \quad
  \uparrow\,\uparrow \,\vert\, \uparrow\,\uparrow, \\
  \mathrm{AFM\text{-}1} &: \quad
  \uparrow\,\downarrow \,\vert\, \uparrow\,\downarrow, \\
  \mathrm{AFM\text{-}2} &: \quad
  \uparrow\,\uparrow \,\vert\, \downarrow\,\downarrow,
\end{align*}
where the vertical bar separates the two original unit cells. In a
classical Ising approximation to the Heisenberg model, each bond
contributes an energy $J_{ij}\,\langle \mathbf{S}_i\!\cdot\!\mathbf{S}_j
\rangle$, with $\langle \mathbf{S}_i\!\cdot\!\mathbf{S}_j\rangle =
+\tfrac{1}{4}$ for parallel spins and $-\tfrac{1}{4}$ for antiparallel
spins. Writing the DFT energies for the three configurations as
$E_{\mathrm{FM}}$, $E_{\mathrm{AFM\text{-}1}}$, and
$E_{\mathrm{AFM\text{-}2}}$, and counting how many $J_1$ and $J_2$ bonds
are parallel or antiparallel in each configuration yields a linear system
of the form
\begin{align}
  E_{\mathrm{FM}} &= E_0 + a_1 J_1 + a_2 J_2, \\
  E_{\mathrm{AFM\text{-}1}} &= E_0 + b_1 J_1 + b_2 J_2, \\
  E_{\mathrm{AFM\text{-}2}} &= E_0 + c_1 J_1 + c_2 J_2,
\end{align}
where $E_0$ is a spin-independent reference energy and the coefficients
$a_i,b_i,c_i$ are simple rational numbers proportional to $\pm\tfrac{1}{4}$.
Eliminating $E_0$ and solving for $J_1$ and $J_2$ gives the desired
exchange constants.

In our nanographene spin-$1/2$ chain, this mapping yields a very weak
$J_1$ and a much larger $J_2$, i.e.\ $J_1 \simeq 0$ and $J_2>0$, placing
the system extremely close to the limit of isolated dimers. This is the
parameter regime used in the ED calculations for the “odd-Haldane”
spin-$1/2$ phase.

\subsection*{S2.4 Mapping collinear DFT+$U$ energies to Hund
exchange and inter-monomer $J_1,J_2$: spin-1 chain}

For the effective spin-1 chain, the building blocks are Hund-coupled
superatoms: each nanographene monomer hosts two spin-$\tfrac{1}{2}$
sites, coupled ferromagnetically by a large intra-monomer exchange
$J_1<0$, while neighboring monomers are coupled antiferromagnetically
by $J_2>0$. In the spin-1 description these correspond to the
intra- and inter-monomer bonds of the Heisenberg chain in Eq.~(1).

In practice we do not extract $J_1$ and $J_2$ from separate “isolated
monomer” and “two-monomer” calculations. Instead we use exactly the
same four-site mapping as for the spin-$\tfrac{1}{2}$ alternating chain,
but now the four spin-$\tfrac{1}{2}$ sites belong to a single
crystallographic unit cell containing two monomers,
\[
  A \equiv (1,2), \qquad B \equiv (3,4),
\]
with $J_1$ bonds inside each monomer (1–2 and 3–4) and $J_2$ bonds
connecting the monomer $A$ to monomer $B$ along the
backbone. All symmetry-equivalent $J_2$ bonds within the unit cell are
assumed equal.

We evaluate DFT+$U$ total energies for the following three collinear
spin configurations of the four sites,
\begin{align*}
  \mathrm{FM} &: \quad
  \uparrow\,\uparrow \,\vert\, \uparrow\,\uparrow, \\
  \mathrm{AFM\text{-}1} &: \quad
  \uparrow\,\downarrow \,\vert\, \uparrow\,\downarrow, \\
  \mathrm{AFM\text{-}2} &: \quad
  \uparrow\,\uparrow \,\vert\, \downarrow\,\downarrow,
\end{align*}
where the vertical bar again separates monomer $A$ from monomer $B$.
In each pattern, every $J_1$ and $J_2$ bond is either parallel or
antiparallel, so its contribution to the Heisenberg energy can be
written as $J\,\langle \mathbf{S}_i\!\cdot\!\mathbf{S}_j\rangle$ with
$\langle \mathbf{S}_i\!\cdot\!\mathbf{S}_j\rangle = +\tfrac{1}{4}$ for
parallel spins and $-\tfrac{1}{4}$ for antiparallel spins.

Denoting the DFT+$U$ total energies of these three configurations by
$E_{\mathrm{FM}}$, $E_{\mathrm{AFM\text{-}1}}$ and
$E_{\mathrm{AFM\text{-}2}}$, and counting how many $J_1$ and $J_2$ bonds
are parallel or antiparallel in each case, we obtain the same
algebraic mapping as in Sec.~S2.3 for the spin-$\tfrac{1}{2}$ chain: the
three equations are sufficient to eliminate $E_0$ and solve uniquely for
$J_1$ and $J_2$.

The crucial difference lies in the resulting parameter regime and its
physical interpretation. For the spin-$\tfrac{1}{2}$ chain we obtain two
antiferromagnetic couplings $J_1,J_2>0$ with $J_1\simeq 0$ and $J_2>0$,
corresponding to an almost perfectly dimerized AFM chain. For the
Hund-coupled spin-1 chain, applying the same four-site mapping to the
unit cell yields
\[
  J_1 < 0, \qquad J_2 > 0, \qquad |J_1| \gg J_2,
\]
i.e.\ a large ferromagnetic intra-monomer Hund exchange $J_1$ that locks
the two spin-$\tfrac{1}{2}$ sites within each monomer into an effective
$S=1$ superatom, and a weaker antiferromagnetic inter-monomer exchange
$J_2$ that couples neighboring $S=1$ superatoms along the chain.  In the notation of the main text we therefore identify
\[
  J_1 = J_{\mathrm{FM}} \quad \text{(intra-monomer Hund exchange)}, \qquad
  J_2 = J_{\mathrm{AFM}} \equiv J_{\mathrm{eff}} \quad
  \text{(inter-monomer AFM exchange)}.
\] 
These
$J_1$ and $J_2$ values are the ones used as input for the ED and DMRG
calculations of the interacting Haldane spin-1 chain in the main text.

\section*{S3. Finite-Temperature Molecular Dynamics and Time-Averaged Structures}

\subsection*{S3.1 MD setup}

Classical molecular dynamics (MD) was performed with \textsc{LAMMPS} \cite{Plimpton1995} using the fine-tuned SevenNet-MF-OMPA machine-learning interatomic potentials.\cite{Kim2025,Park2024, Anindya2025} The simulation cell is quasi-one-dimensional with periodic boundary conditions only along the chain direction (\texttt{boundary p f f}). The 0~K relaxed structure served as the starting geometry. All simulations employed metal units, a time step of \(\Delta t = 0.0005~\mathrm{ps}\) (0.5~fs), velocity-Verlet integration, a neighbor skin of 2.0~\AA\ rebuilt every step, and linear center-of-mass momentum removal every 100 steps. Parallel domain decomposition was aligned with the chain axis and dynamically load-balanced.

\subsection*{S3.2 Stage I: anisotropic barostat (NPT-$x$) to obtain the stress-free repeat length \(L_x^{\ast}(T)\)}

At a target temperature \(T\), we relaxed only the periodic lattice length using an anisotropic Nosé–Hoover thermostat/barostat\cite{Nose1984,Hoover1985,Martyna1992}:

The thermostat damping time was \(0.2~\mathrm{ps}\) and the barostat damping time along \(x\) was \(8.0~\mathrm{ps}\), with additional barostat drag (1.5) to suppress ringing. The total NPT-$x$ duration was \(t_{\mathrm{NPT}}=40~\mathrm{ps}\), split into a \emph{settle} window of \(25~\mathrm{ps}\) (not averaged) followed by an \emph{average} window of \(15~\mathrm{ps}\). Over the average window we recorded the scalar averages
\begin{align}
L_x^{\ast}(T) &= \big\langle L_x \big\rangle \equiv \frac{1}{N_{\mathrm{avg}}}\sum_{k=1}^{N_{\mathrm{avg}}} L_x(k),\\
\sigma(L_x) &= \sqrt{\big\langle L_x^2 \big\rangle - \big\langle L_x \big\rangle^2},\qquad
\big\langle P_{xx} \big\rangle = \frac{1}{N_{\mathrm{avg}}}\sum_{k=1}^{N_{\mathrm{avg}}} P_{xx}(k),
\end{align}
using time-averaging fixes. Because the cell is non-periodic in \(y,z\), instantaneous \(P_{xx}\) exhibits large fluctuations; the zero-tension criterion is therefore based on the long-time mean \(\langle P_{xx}\rangle\) together with the observed plateau of \(L_x(t)\). After NPT-$x$, the box length was set to the averaged value \(L_x^{\ast}(T)\) via a \texttt{change\_box} operation (preserving \(x_{\mathrm{lo}}\)).

\subsection*{S3.3 Stage II: NVT at fixed \(L_x=L_x^{\ast}(T)\) and time-averaged geometry}

With \(L_x\) fixed to \(L_x^{\ast}(T)\), we ran a Nosé–Hoover NVT at the same temperature:

The NVT stage comprised a \(10~\mathrm{ps}\) settle segment followed by a \(20~\mathrm{ps}\) production segment. During production we accumulated arithmetic mean atomic coordinates with a per-atom time-averaging fix to obtain
\begin{equation}
\overline{\mathbf r}_i=\frac{1}{N_{\mathrm{prod}}}\sum_{k=1}^{N_{\mathrm{prod}}} \mathbf r_i(k)\,,
\qquad i=1,\dots,N_{\mathrm{atoms}},
\end{equation}
and monitored \(L_x(t)\) to confirm stability at \(L_x^{\ast}(T)\). A trajectory was written every 1000 steps for distributional analyses (bond/angle/dihedral and transverse fluctuations), and an instantaneous end-of-run snapshot and restart were saved for reproducibility.

\subsection*{S3.4 Thermal expansion and uncertainties}

Repeating the above at several temperatures yields \(L_x^{\ast}(T)\) versus \(T\). The one-dimensional linear thermal expansion coefficient is obtained by a linear fit,
\begin{equation}
\alpha \equiv \frac{1}{L_x^{\ast}(0)}\frac{\mathrm d L_x^{\ast}(T)}{\mathrm d T},
\end{equation}
with uncertainties estimated from the standard errors of \(L_x^{\ast}(T)\) over the NPT-$x$ averaging windows and propagated through the fit.

\medskip
All numerical parameters (damping times, settle/average durations, and dump cadences) match those used in production and were chosen to suppress barostat ringing in the quasi-1D geometry while providing statistically converged late-time averages.



\section*{S4. Finite-Temperature Molecular Dynamics: 300 K Structural Stability}

\subsection*{S4.1 Unified interpretation (applies to both chains)}
Two complementary diagnostics summarize the 300 K structural response under periodicity along the chain axis. 
(i) The end-to-end trace (minimum-image disabled; both termini kept in the same box image) is a stability check: a narrow, stationary band indicates a stress-balanced repeat length and the absence of creep, progressive buckling, or slow uncoiling on the simulated timescale. 
(ii) The transverse RMS profile provides a flexibility map: peaks at inter-monomer linkages and substituents identify hinge-like regions, whereas valleys on aromatic carbons mark a stiff conjugated backbone. Out-of-plane fluctuations remain near the numerical floor, consistent with an essentially planar geometry at 300 K.

\subsection*{S4.2 Chain-specific notes}
\textbf{Spin-\(\tfrac{1}{2}\) chain.} The end-to-end distance forms a tight, stationary band (mean \(\approx 16.13\) Å, s.d. \(\approx 0.10\) Å over the production window). The transverse RMS highlights localized flexibility at junctions and side groups while the aromatic cores remain rigid in-plane.  
\textbf{Spin-1 chain.} The end-to-end trace is likewise stationary with small-amplitude thermal modulations (mean \(\approx 16.14\) Å, s.d. \(\approx 0.04\) Å). The transverse RMS exhibits the same qualitative pattern—enhanced motion at junctions/substituents; near-zero out-of-plane fluctuations—confirming preserved planarity at 300 K.

\subsection*{S4.3 Implications}
Both chains maintain a well-defined, planar backbone at 300 K, with thermal motion concentrated at chemically sensible soft spots. This supports the use of a single, temperature-specific geometry and modest fluctuation envelopes for subsequent electronic and magnetic analyses.

\begin{figure}[h]
  \centering
  \begin{subfigure}[t]{0.49\linewidth}
    \centering
    \includegraphics[width=\linewidth]{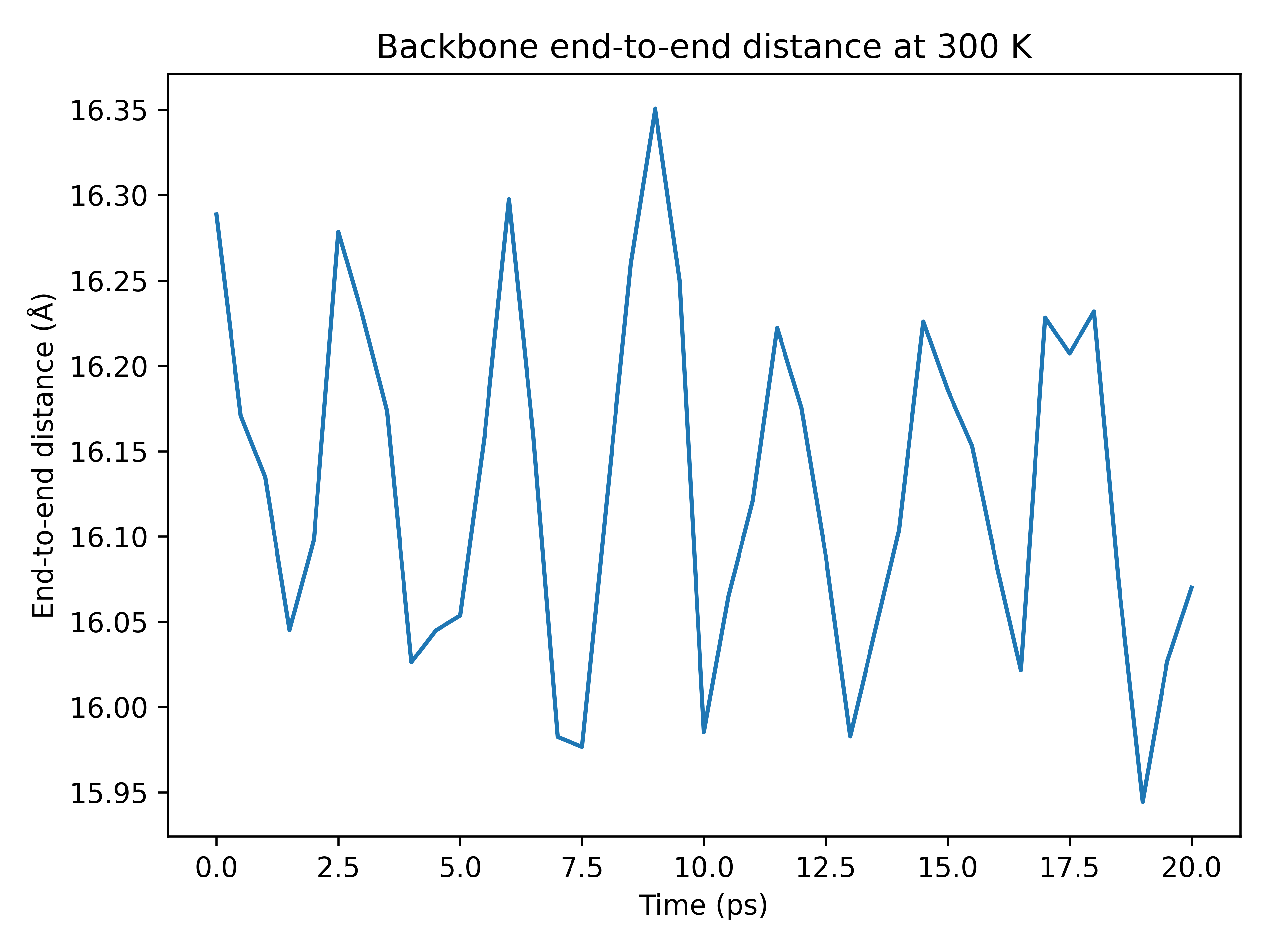}
    \subcaption{Spin-\(\tfrac{1}{2}\): end-to-end distance at 300 K. Stationary, narrowly distributed trace indicating a stress-balanced repeat length.}
  \end{subfigure}
  \hfill
  \begin{subfigure}[t]{0.49\linewidth}
    \centering
    \includegraphics[width=\linewidth]{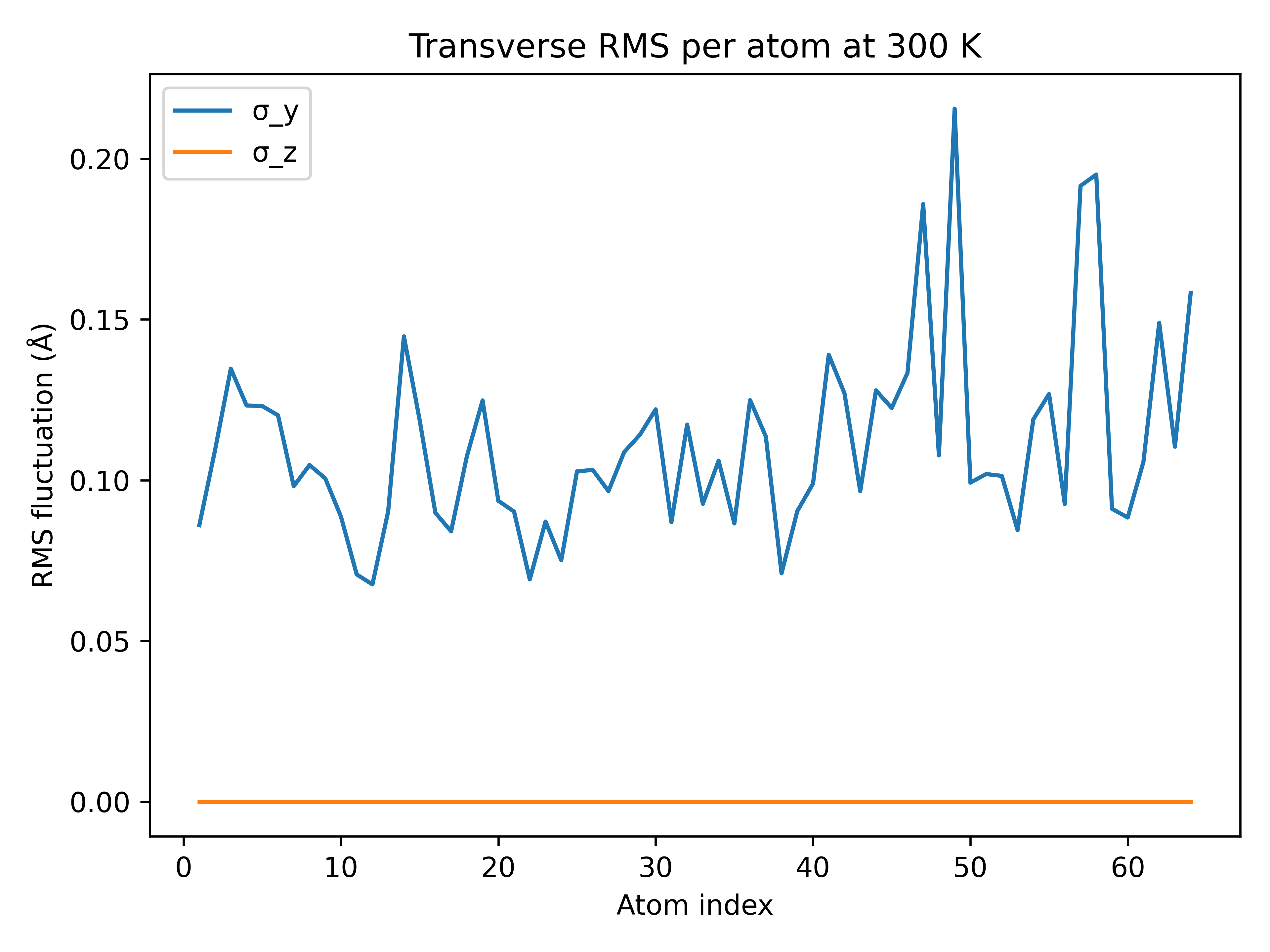}
    \subcaption{Spin-\(\tfrac{1}{2}\): transverse RMS per atom at 300 K. Flexibility localizes at junctions and substituents; the aromatic backbone remains rigid and essentially planar.}
  \end{subfigure}

  \vspace{0.5em}
  \begin{subfigure}[t]{0.49\linewidth}
    \centering
    \includegraphics[width=\linewidth]{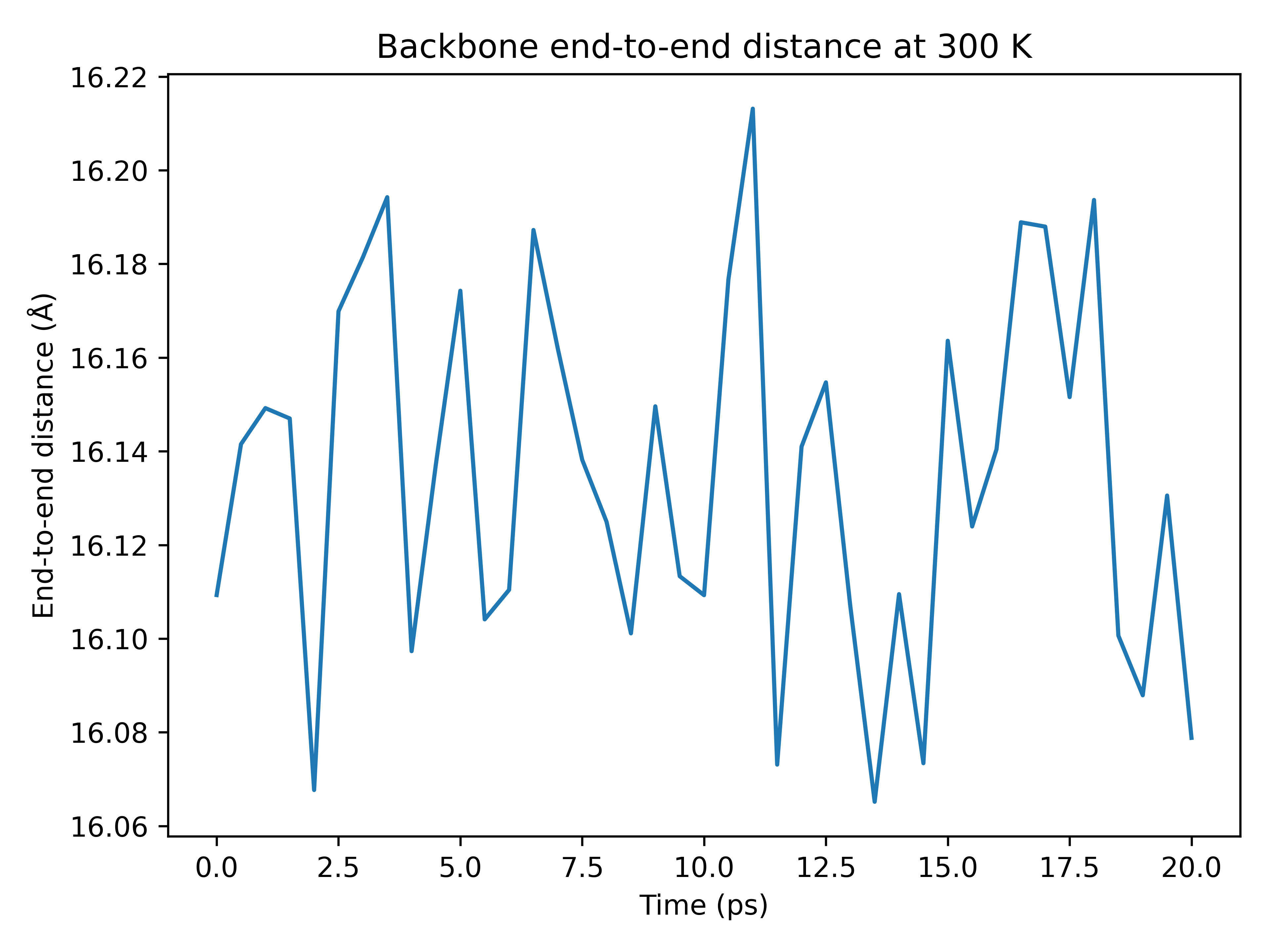}
    \subcaption{Spin-1: end-to-end distance at 300 K. Narrow, drift-free band consistent with a stress-balanced repeat length.}
  \end{subfigure}
  \hfill
  \begin{subfigure}[t]{0.49\linewidth}
    \centering
    \includegraphics[width=\linewidth]{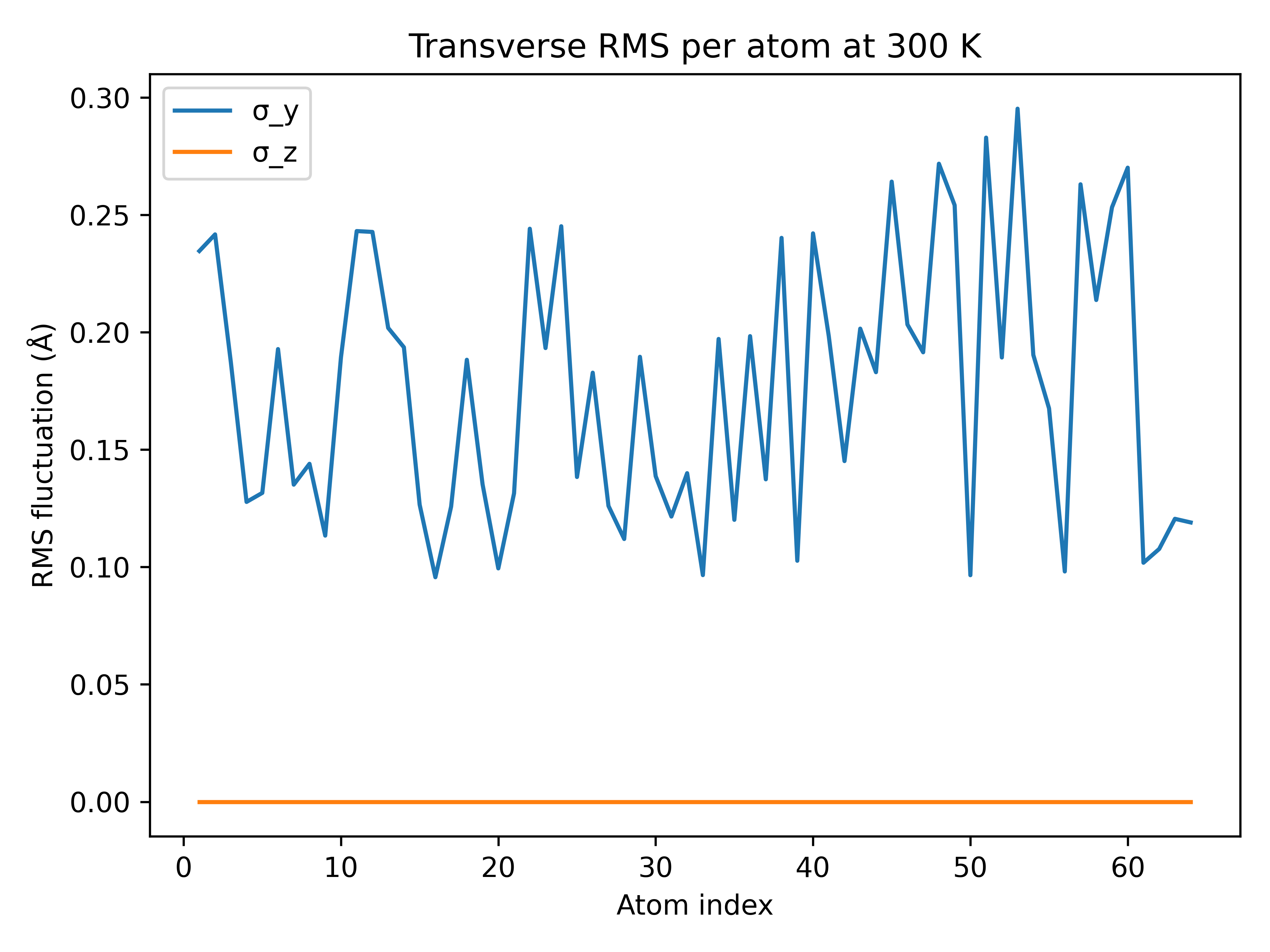}
    \subcaption{Spin-1: transverse RMS per atom at 300 K. Junction-localized flexibility with near-zero out-of-plane motion; backbone remains planar.}
  \end{subfigure}

  \caption{Structural stability diagnostics at 300 K for the spin-\(\tfrac{1}{2}\) and spin-1 chains. End-to-end traces serve as stability checks; transverse RMS profiles map flexibility along the backbone.}
  \label{fig:S4_md_300K}
\end{figure}

\bibliography{acs-latex-template}        



%
%
%
%
%